\documentclass[]{raa}
\usepackage{lscape}
\usepackage[encapsulated]{CJK}
\usepackage{mathrsfs}
\usepackage{epstopdf}
\usepackage{longtable}
\usepackage{setspace}
\usepackage{hyperref} 
\usepackage{threeparttable}
\usepackage{diagbox}
\usepackage{lineno}
\usepackage{graphicx,times}
\usepackage{natbib}
\usepackage{amssymb,amsmath}
\usepackage{color}
\usepackage{float}
\bibpunct{(}{)}{;}{a}{}{,}

\newbox\grsign \setbox\grsign=\hbox{$>$} \newdimen\grdimen \grdimen=\ht\grsign
\newbox\laxbox \newbox\gaxbox
\setbox\gaxbox=\hbox{\raise.5ex\hbox{$>$}\llap
	{\lower.5ex\hbox{$\sim$}}}\ht1=\grdimen\dp1=0pt
\setbox\laxbox=\hbox{\raise.5ex\hbox{$<$}\llap
	{\lower.5ex\hbox{$\sim$}}}\ht2=\grdimen\dp2=0pt

\newcommand{\choh}{CH$_3$OH}
\newcommand{\ho}{H$_2$O}

\begin{document}
	\begin{CJK*}{UTF8}{gbsn}
		
   \title{VLBI with SKA: Possible Arrays and Astrometric Science}

\volnopage{ {\bf 20XX} Vol.\ {\bf X} No. {\bf XX}, 000--000}
\setcounter{page}{1}

\author{Yingjie Li
	\inst{1}, Ye Xu\inst{1, 2}, Jingjing Li\inst{1}, Shuaibo Bian
	\inst{1},  Zehao Lin \inst{1}, Chaojie Hao\inst{1}, Dejian Liu\inst{1, 2}
}

\institute{ Purple Mountain Observatory, Chinese Academy of Sciences, Nanjing 210023, China {\it xuye@pmo.ac.cn, liyj@pmo.ac.cn}\\
	\and
	University of Science and Technology of China, Hefei, Anhui 230026, China\\
	\vs \no
	{\small Received 20XX Month Day; accepted 20XX Month Day}
}   	
		
\abstract{
The next generation of very long baseline interferometry (VLBI) is stepping into the era of microarcsecond ($\mu$as) astronomy, and pushing astronomy, especially astrometry, to new heights. VLBI with the Square Kilometre Array (SKA), SKA-VLBI, will increase current sensitivity by an order of magnitude, and reach astrometric precision routinely below 10 $\mu$as, even challenging 1 $\mu$as. This advancement allows precise parallax and proper motion measurements of various celestial objects. Such improvements can be used to study objects (including isolated objects, and binary or multiple systems) in different stellar stages (such as star formation, main-sequence stars, asymptotic giant branch stars, pulsars, black holes, white dwarfs, etc.), unveil the structure and evolution of complex systems (such as the Milky Way), benchmark the international celestial reference frame, and reveal cosmic expansion. Furthermore, the theory of general relativity can also be tested with SKA-VLBI using precise measurements of light deflection under the gravitational fields of different solar system objects and the perihelion precession of solar system objects.			
}

\authorrunning{Y. J. Li et al. }            
\titlerunning{SKA-VLBI Astrometry}  
\maketitle

\section{Introduction}\label{sec:intro}

Astrometry focuses on how to measure the positions and motions of celestial objects as accurately as possible. The history of astrometry originates from the earliest astronomical observations recorded more than 2000 years ago. Successive improvements in astrometric technology have increased the angular accuracy, leading to a number of important discoveries, such as Earth nutation and the proper motions, aberrations, and trigonometric parallaxes of stars. Such improvements have enhanced marine and aerospace navigation after the international celestial reference frame (ICRF) was established, based on accurate positional measurements of a large number of celestial bodies \citep{Perryman2012}. The application of astrometry has continuously extended, from the solar system, to the whole Milky Way (MW), to galaxies, and the entire visible universe; the development of theories that depend on precise astrometry, such as heliocentric theory, Newtonian spacetime, and the theory of general relativity (GR), has occurred concurrently.

The history of astrometry is closely associated with that of stellar catalogs, because they offer reference points on the sky for astronomers to trace the motion of stars. In 1830s, the first trigonometric parallax of a star was measured after many unsuccessful attempts \citep{Bessel1838}, which opened up a new era of trigonometric parallax measurements. Over the past 400 years, the accuracy of positional measurements has increased almost logarithmically with time. This increase of accuracy has allowed astronomers to quantify the extremely large scale of the universe, determine the physical properties of stars, and ultimately describe the structure, dynamics, and the origin of our MW galaxy \citep[for more details about the history of astrometry see][]{Perryman2012}.

Astronomical observations in the radio started in the 1920s with Karl Jansky, and was further driven by the development of radar during World War II \citep{Reid-Honma2014}. The earliest realization of applying interferometric techniques to practical astronomical observation dates back to the 1920s, when the first two-element optical interferometer was used \citep{Thompson+2017}. The technique of two-element interferometry was applied to astronomical radio observations in the 1940s \citep{Ryle-Vonberg1946}, and the realization of voltage multiplier correlators dates back to 1952 \citep{Thompson+2017}.

Since the middle of the 20th century, radio and space optical astrometry has developed side by side, providing mutual reinforcement, ushering in breakthroughs in astrometry. In the 1950s--1960s, radio interferometric techniques continuously improved, increasing astrometric accuracy continuously from $\sim$1$''$ to 0.03$''$ \citep{Wade-Johnston1977}. In the late 1960s, the use of precise atomic clocks and magnetic tapes to record data made it possible to overcome baseline length limitations by not having to transfer radio signals directly through cables, which eventually revolutionized radio interferometry. This new technology was dubbed very long baseline interferometry \citep[VLBI;][]{Reid-Honma2014}, and the first successful radio VLBI was in 1967 \citep{Thompson+2017}. During the middle of the 1970s, different facilities were tied together to form arrays with more than 10 elements. In this early stage of VLBI, the absolute positional accuracy was better than $\sim$0.3 mas \citep{Clark+1976, Ma+1986}, and relative astrometric accuracy between fortuitously close pairings of quasi-stellar objects (QSOs) reached $\sim$10 $\mu$as \citep{Marcaide+1985}.

With improvements in positional precision, GR effects gradually became obvious and affected the precision of astrometry. For instance, from GR predictions, light should be deflected by $\sim$500 mas if it passes within 1$^{\circ}$ of the Sun, and the maximum deflection angle caused by Jupiter exceeds $\sim$16 mas \citep[e.g.,][]{Li+2022}. In the 1980s, GR models for light propagation began to be applied to VLBI observations, and a ``consensus model'' was published in 1991 \citep[see][]{Eubanks1991, Pertit-Luzum2010}. Currently, the relative time delays between antenna pairs caused by the Sun, the planets, the
Moon, and the large satellites of Jupiter, Saturn, and Neptune
are determined using data correlators, which are equivalent to deducting the deflection angles caused by the corresponding lenses \citep[see, e.g.,][]{Pertit-Luzum2010}. The use of data correlators has further enhanced high-precision astrometry.

In the same year as the first successful radio VLBI observation, it was proposed that precise measurements of triangular parallax from space could be achieved. This assertion was made a reality with the launch of the optical satellite Hipparcos (launched in August 1989 and ceased operating in March 1993). Before the middle of the 1990s, the number of stars that had parallax measurements made by ground facilities was only $\sim$8000 \citep{van-Altena+1995}. The Hipparcos satellite provided the positions, parallaxes, and proper motions of approximately 120,000 stars with high accuracy \citep[i.e., up to $\sim$1 mas,][aulthough this accuracy was inferior to the accuracy of radio VLBI measurements]{ESA1997}. Later catalogs included the Tycho 2 catalog \citep[see][]{Hog+2000} containing 2.5 million stars with improved astrometric quality, albeit mainly for the brightest stars \citep{van-Leeuwen2010}. Ever since radio VLBI and the Hipparcos satellite determined the positions of celestial objects, comparisons between optical and radio astrometry have been carried out. The most prominent example is the comparison between the optical and radio reference frames. Eventually, the VLBI-based ICRF \citep[i.e., ICRF1,][]{Ma+1998} was adopted by the International Astronomical Union (IAU) in 1995. In this era, the controversy existed between the optical and radio reference frames, and there was more to it than that, such as the famous ``Pleiades distance controversy'' \citep{Melis+2014}. In this controversy, the parallax obtained by the Hipparcos satellite was different from the distance obtained by other methods, including parallax measurement by VLBI. It has been described as ``one of the more dramatic controversies in modern astrophysics'' \citep{Paczynski2003}.\footnote{See \url{https://www.cosmos.esa.int/web/hipparcos/pleiades-distance}.}

The first array of antennas built specifically for astronomical VLBI measurements
was the Very Long Baseline Array, VLBA, of the National Radio
Astronomy Observatory (NRAO),\footnote{The National Radio Astronomy Observatory is a facility of the US National Science Foundation operated under cooperative agreement by Associated Universities, Inc.} which began operations in 1994 \citep{Thompson+2017}. A VLBI dedicated to astrometry, and more specifically, to obtain a 3-dimensional (3D) map of the MW galaxy, is the Japanese VLBI Exploration of Radio Astrometry (VERA), which has been under regular operation since the fall of 2003.\footnote{See \url{https://www.miz.nao.ac.jp/veraserver/}.\label{web:VERA}} Stepping into the 2000s, with the improvement of calibration techniques used in VLBI, a relative position accuracy of $\sim$10 $\mu$as was routinely achieved for most bright targets \citep{Reid-Honma2014}, including a record parallax precision of $\pm$ 5 $\mu$as \citep{Zhang+2013}. Contemporaneously, Gaia, as the successor to the Hipparcos space astrometry project, launched in 2013. To date, there have been three data releases from Gaia \citep[i.e., DR1, DR2, and (E)DR3;][]{Gaia-Collaboration+2016, Gaia-Collaboration+2021, Gaia-Collaboration-DR3-2022}. At present, Gaia's astrometric accuracy is 20--30 $\mu$as, and is expected to reach $\sim$7 $\mu$as in data release 5. The first priority of Gaia was to address the ``Pleiades distance controversy'' \citep{Gaia-Collaboration+2016}. Gaia DR2 successfully ended the controversy, finding a distance consistent with that obtained by other methods including VLBI triangular parallax measurements \citep[a summary of distance measuments is given in][]{Lodieu+2019}.

VLBI and Gaia astrometry have had a profound effect on many fields in astronomy. Two important areas of comparative research are the comparison of the optical and radio reference frames and the study of the structure of the MW, which are still active today. In the long-term comparison of the optical and radio reference frames, notable VLBI-Gaia positional offsets, i.e., offsets between measured optical and radio positions, have been discovered. Specifically, these offsets are found in the positions in the most recent, third release of the ICRF, \citep[ICRF3;][]{Charlot2020}, and from the Gaia Celestial Reference Frame (GCRF) based on EDR3 \citep{Gaia-Collaboration+2022}. The precision of Gaia astrometry is constantly improving, and solving these VLBI-Gaia positional offsets also requires a continual improvement of radio VLBI positional precision and spatial resolution as well as multiband radio and multiple calibrator observations.

The structure of the MW cannot be studied by the data produced by Hipparcos, because the distances of stars that could be accurately determined was limited to $\sim $ 100 pc in the Tycho-2 star catalog \citep{ESA1997}. The accurate measurement of the distance to the high-mass star-forming region (HMSFR) W3(OH) using VLBA \citep[with a parallax accuracy up to 0.01 mas;][]{Xu+2006}, marked the possibility of making direct measurements of the spiral arm structure of Galaxy, opening a new era in the study of the structure of the MW. At present, precise triangular parallaxes have been measured for over 200 masers associated with HMSFRs in the Bar and Spiral Structure Legacy (BeSSeL) survey\footnote{See \url{http://bessel.vlbi-astrometry.org}.} \citep{Brunthaler+2011} and the VERA project\textsuperscript{\ref{web:VERA}} \citep{VERA-Collaboration+2020}. These surveys have achieved fruitful results, such as mapping the nearest spiral arm, the Local arm, in unprecedented detail \citep{Xu+2013, Xu+2016}, generating the most complete spiral arm structure pattern of the MW to date \citep{Reid+2019}. Data from Gaia can densify the pattern of the spiral structure of the MW near the Sun (within $\sim$5 kpc) and even allow investigations of the evolution of its spiral structure. For instance, the structures traced by young OB stars are consistent with those traced by masers associated with HMSFRs \citep[e.g.,][]{Xu+2018raa, Xu+2018, Xu+2021}. Additionally, the large age spans of open clusters made them suitable for studying the motion and evolution of spiral arms \cite[e.g.,][]{Hao+2021}, and even main-sequence stars and old stars can be used to constrain spiral arm structure \cite[e.g.,][]{Poggio+2021, Lin+2022}.

The comprehensive utilization of high-precision astrometric data from VLBI and Gaia has promoted research on the structure and evolution of the Milky Way. But there are still some issues that need to be solved. For example, using high-precision data of various young stellar objects (including masers, OB stars and young open clusters), \citet{Xu+2023} proposed a new model of Galactic spiral arm that appears to be different from the previous mainstream view depicted by \citet{Georgelin-Georgelin1976} and \citet{Reid+2019}. However, there is still a lack of sufficient data to depict the spiral structure across the southern sky and beyond the Galactic center (GC). Will the spiral arm structure traced by masers still be consistent with that traced by OB stars \citep[see][]{Xu+2021}? Is the formation of stars in the spiral arms uneven \citep{Xu+2021}? These questions may be answerable with the next generation of radio telescopes.

The next generation of instruments, such as the Square Kilometre Array \citep[SKA; including SKA-Mid and SKA-Low;][]{Braun+2015} will greatly improve sensitivity (by an order of magnitude) and cover a wide range of frequencies from dozens of MHz to dozens of GHz. The concept of VLBI with the SKA (SKA-VLBI) was introduced for the first time in \citet{Paragi+2015}. SKA-VLBI is about phasing up the core of SKA telescopes, providing multiple tied-array beams that would target a number of sources within the primary field of view (FoV) of the antennas that form a VLBI array together with the SKA. The Karl G. Jansky Very Large Array (JVLA) in the USA (as part of the High Sensitivity Array, HSA) and the Westerbork Synthesis Radio Telescope in the Netherlands (formerly part of the European VLBI network, EVN) provided a single tied array beam in the past; what is new in SKA-VLBI is that it adds a very sensitive component to VLBI networks with large-FoV and multi-beam capability.

The large-FoV will enable SKA-VLBI to offer the multiple in-beam sources astrometry approach \citep{Fomalont-Reid2004}. This has been done with different arrays (VLBA, EVN), albeit typically using only one calibrator. The great benefit is that it allows for simultaneous observation of the target source and calibrators. And the major improvement in precision can be achieved with a number of in-beam sources that individually not, but when their signal is combined they do reach dynamic ranges (DR) of $\sim$100:1. Another advanced technology is using multiple bright calibrators in the sky at several degrees apart, and fit for a phase-screen across the sky, which was called MultiView \citep{Fomalont-Kopeikin2003, Fomalont2005, Rioja-Dodson2020}. Both methods will already help removing the bulk of the systematics, although it is true the final astrometric precision will depend on various factors (like the brightness, compactness, the number, and the distribution of these calibrators) and may reach routinely below 10 $\mu$as, even challenging 1~$\mu$as. 
It should be noted that extremely high-precision astrometry (e.g. SKA $\mu$as astrometry) will face the challenge of light deflection under the gravitational fields of celestial bodies in the solar system \citep[e.g.,][]{Li-Xu-Bian+2022}. In any case, there is now a realistic possibility of carrying out $\mu$as radio astrometric surveys of exquisite detail large samples of objects, which will advance our understanding of different astrophysical phenomena and classes \citep{Rioja-Dodson2020}.

\citet{Paragi+2015} described motivation for and the possible realization of SKA-VLBI. This paper would give an extensive review of possible astrometric science with SKA-VLBI. The remainder of this paper is organized as follows. Section \ref{sec:array} describes which telescopes/sites can be used to make up VLBI arrays or networks with SKA-Mid, and their corresponding sensitivities and astrometric precision. Due to a variety of target sources, it is difficult to cover all the scientific subjects that can be carried out in a single study. However, we use a point to point narrative to gradually depict SKA-VLBI astrometric science in Section \ref{sec:science}. We divide the achievable science into astrometry of isolated objects in Section \ref{sec:science-isolate}; of binary and complex systems in Sections \ref{sec:science-binary-Multi} and \ref{sec:science-complex}; and tests of GR in Section \ref{sec:science-GR}. Finally, Section \ref{sec:sum} gives a summary.

\section{Possible VLBI Networks for SKA-Mid} \label{sec:array}

\subsection{The Compositions of Possible VLBI Networks} \label{sec:array-comp}

The main focus of this work is astrometry, and the lowest frequency considered in this work is $\sim$1 GHz. Due to the frequency of te three priority bands currently planned for deployment for SKA-Mid (950--1760, 4600--8500, and 8300--15,300 MHz),\footnote{see \url{https://skao.canto.global/s/MH29C?viewIndex=0}.} the frequency ranges considered in this work are 1.0--2.0 GHz and 4.6--15.3 GHz. They are further divided into eight bands, i.e., bands L1, L2, C1, C2, C3, X, Ku1, and Ku2 centered at 1.3, 1.6, 5.0, 6.0, 6.7, 8.0, 12.0, and 15.0 GHz, respectively.

\begin{figure}
	\centering
	\hspace{-1cm}
	\includegraphics[width=1.5\textwidth, angle = 90]{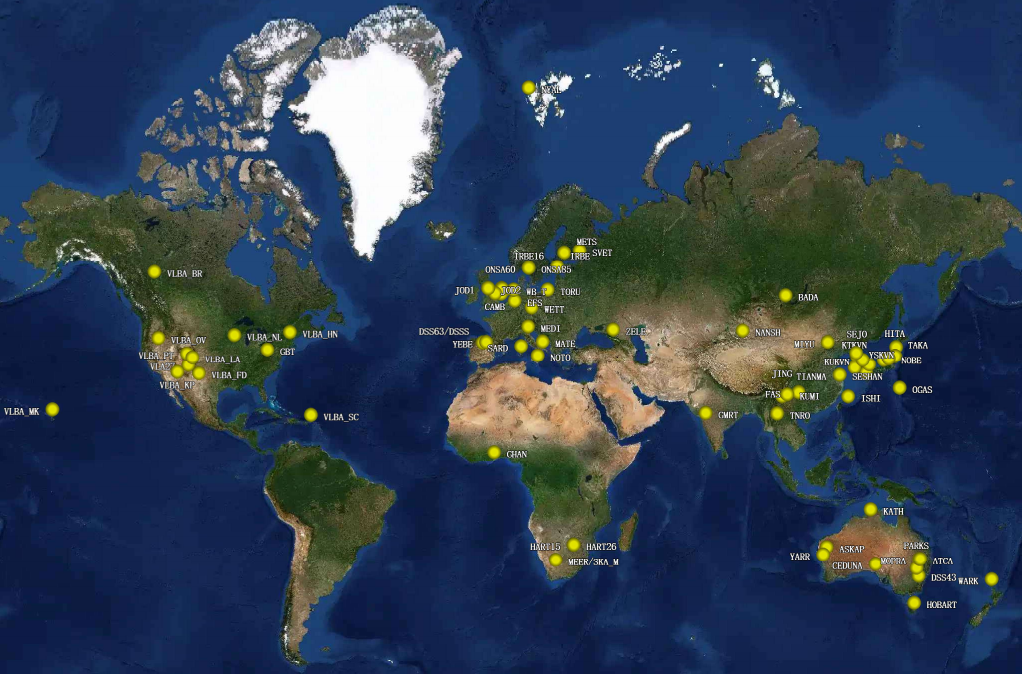}
	\caption{The sites in this work, plotted using ovital map (see \url{https://www.ovital.com/}). SKA-Mid is labeled as ``SKA\_M''.}
	\label{fig:map}
\end{figure}

\begin{figure}
	\centering
	\includegraphics[width=0.9\textwidth]{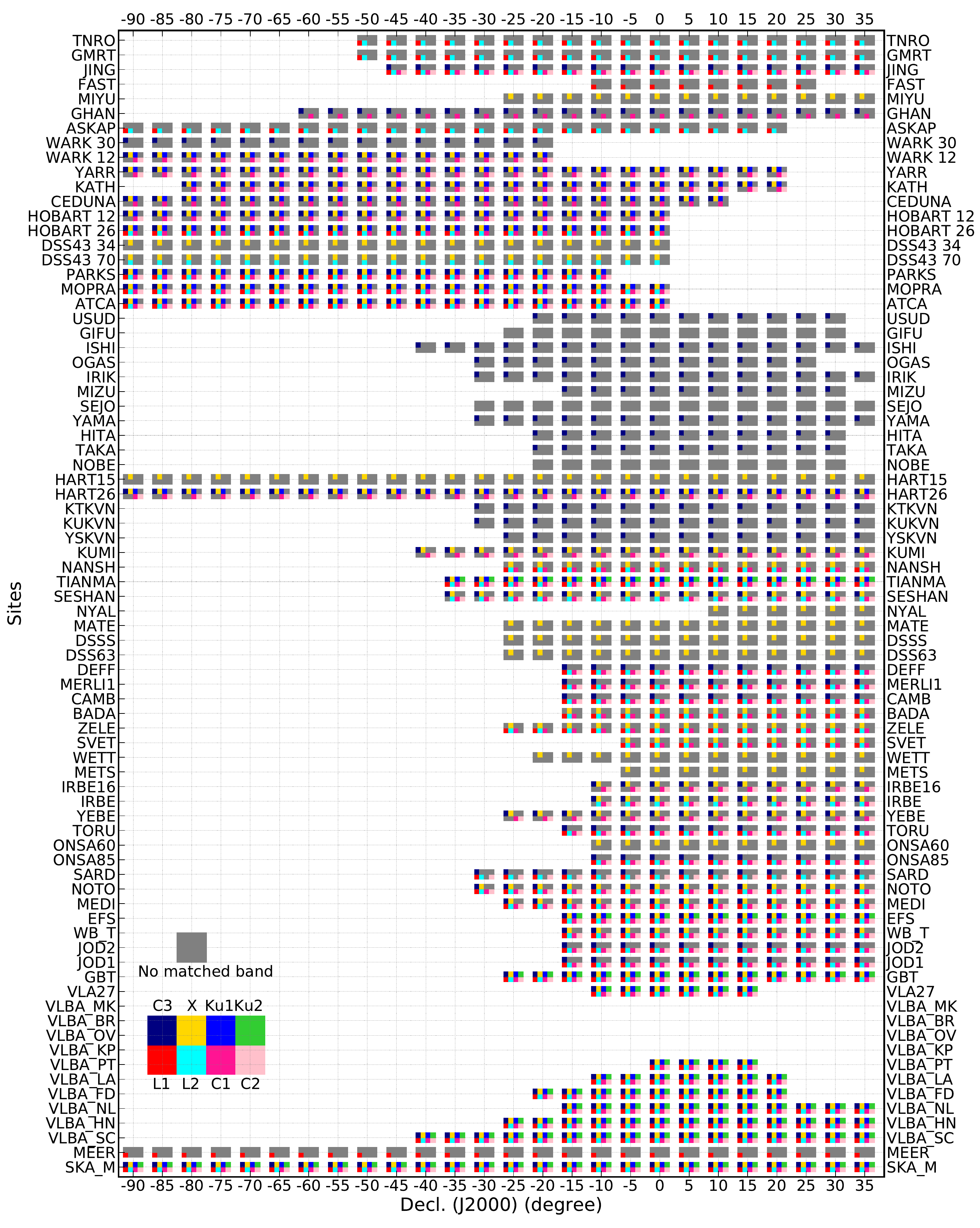}
	\caption{The potential sites of SKA-VLBI for target sources with different Decl. (J2000), where L1, L2, C1, C2, C3, X, Ku1, and Ku2 represent different bands. SKA-Mid is labeled as ``SKA\_M''.}
	\label{fig:conf}
\end{figure}

Currently, the common radio interferometry arrays or networks operating at $\sim$1--15 GHz include the VLBA and JVLA in the USA, the EVN, the East Asia VLBI Network (EAVN), the Australian Long Baseline Array (LBA), and the upgraded Giant Metrewave Radio Telescope (uGMRT) in India, etc. Some subarrays of these arrays or networks (which can work independently) include the Multi-Element Radio Interferometer Network (MERLIN) in the UK, the Korean VLBI Network (KVN), the Japanese VLBI Network (JVN), VERA, and the Chinese VLBI Network (CVN), etc. Large single dishes include the Five-hundred-meter Aperture Spherical radio Telescope (FAST) in China, the Green Bank Telescope (GBT) in the USA, and Effelsberg (EFS) in Germany, etc. Telescope arrays or single dishes under development or construction include the SKA (including SKA-Mid and SKA-Low), next-generation Very Large Array (ngVLA), the African VLBI network (AVN), and Jingdong in China, etc. The total number of telescopes and/or tied arrays involved in this work is 82 (their distribution can be seen in Figure \ref{fig:map} updated on 2023 April 1). These facilities have the potential to be used in VLBI arrays or networks with SKA-Mid. Table \ref{tab:telescopes} lists the basic parameters of these telescopes, such as their geographical position, aperture, and System Equivalent Flux density (SEFD). Figure \ref{fig:conf} shows which telescopes or arrays may be able to form VLBI arrays or networks with SKA-Mid in various bands.

\begin{figure}
	\centering
	\includegraphics[width=0.9\textwidth]{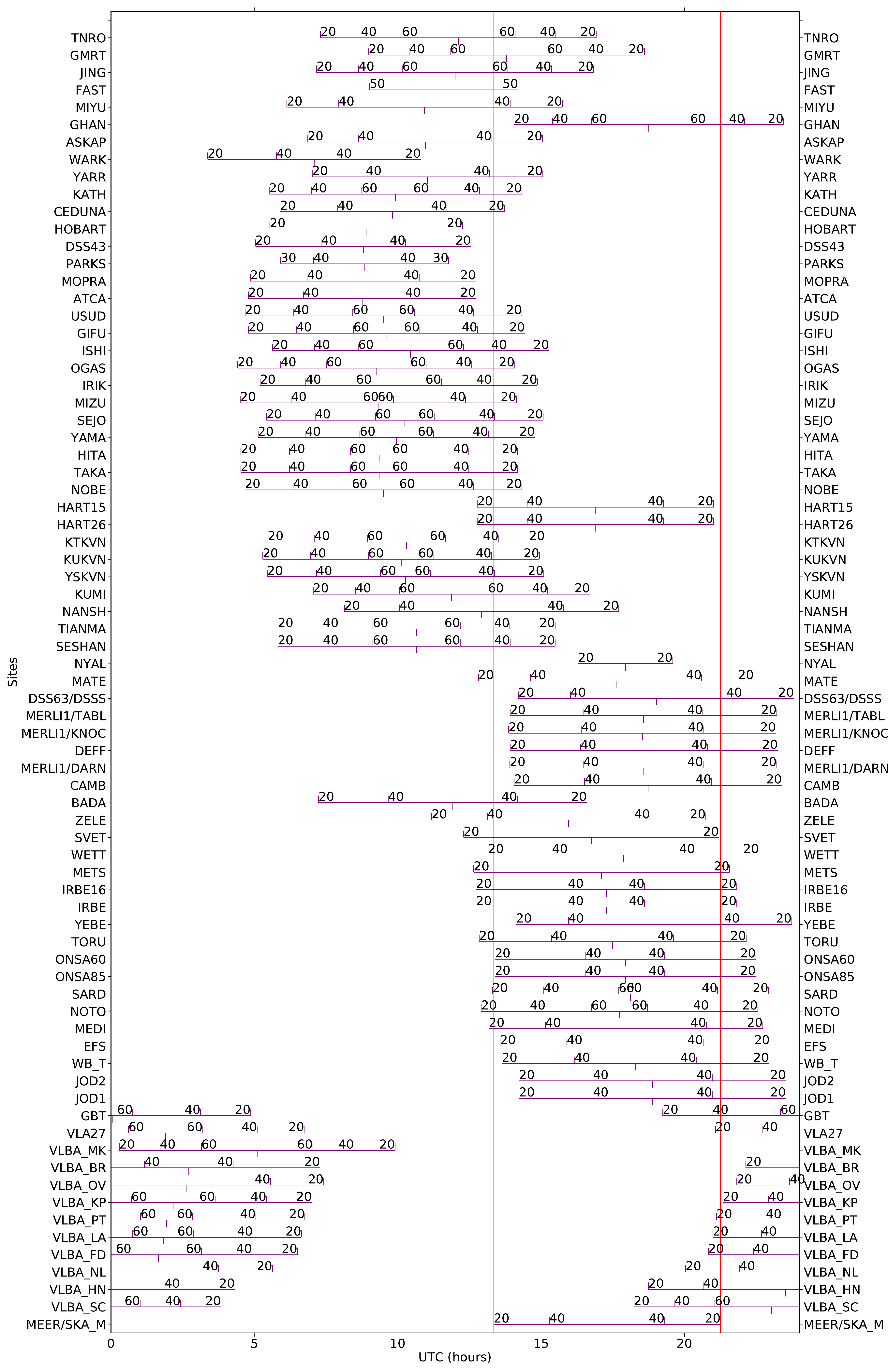}
	\caption{The distribution of elevation for target sources with Decl. (J2000) of 10$^{\circ}$. The red vertical lines indicate the UTC time range with the elevation of SKA-Mid (which is labeled as ``SKA\_M'') being 20$^{\circ}$.}
	\label{fig:elevation}
\end{figure}

\begin{figure}
	\centering
	\begin{subfigure}{0.8\textwidth}
		\includegraphics[width=0.8\textwidth, trim = 0 0 70 15, clip]{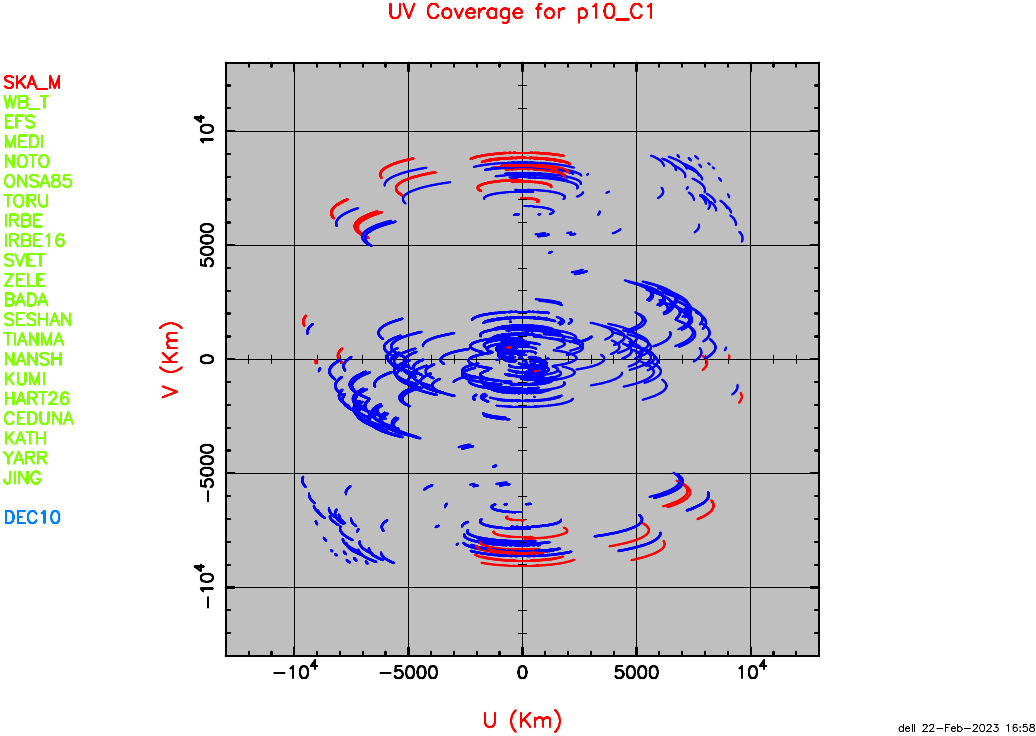}
		\caption{Option 1.}
	\end{subfigure}
	\begin{subfigure}{0.8\textwidth}
		\includegraphics[width=0.8\textwidth, trim = 0 0 70 15, clip]{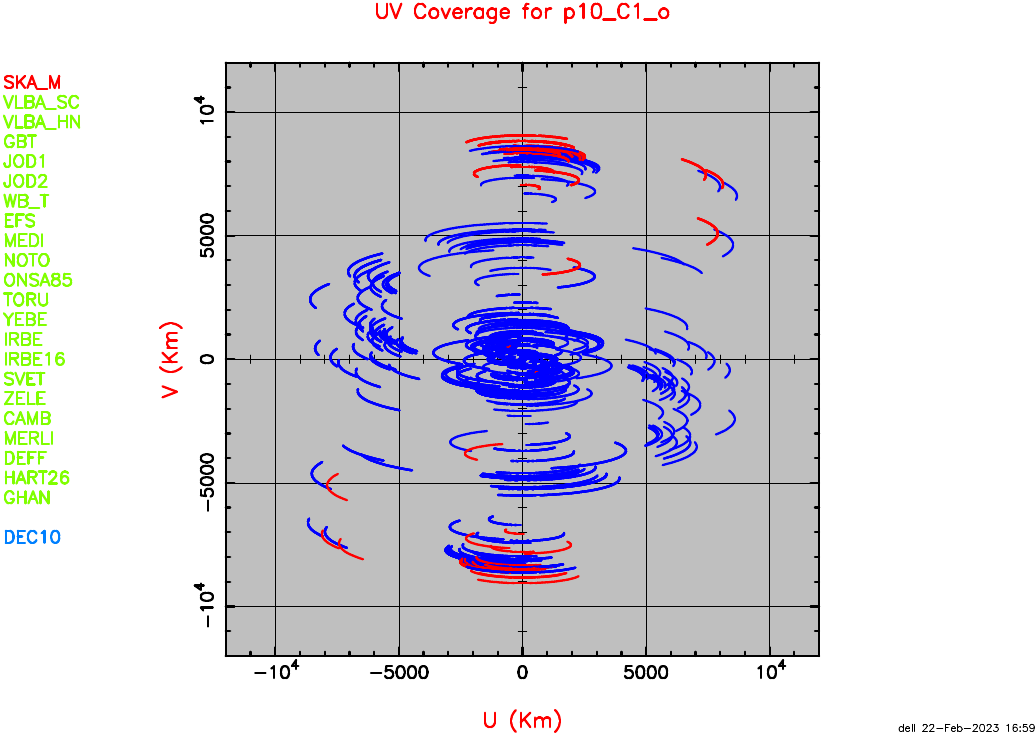}
		\caption{Option 2.}
	\end{subfigure}
	\caption{Map of uv coverage in band C1 for options 1 (panel (a)) and 2 (panel (b)), with target sources' Decl. (J2000) = $10^{\circ}$. The uv coverage was mapped with the tool provided at \url{http://www.aoc.nrao.edu/software/sched/}.}
	\label{fig:uv8}
\end{figure}

\begin{figure}
	\centering
    \includegraphics[width=0.8\textwidth]{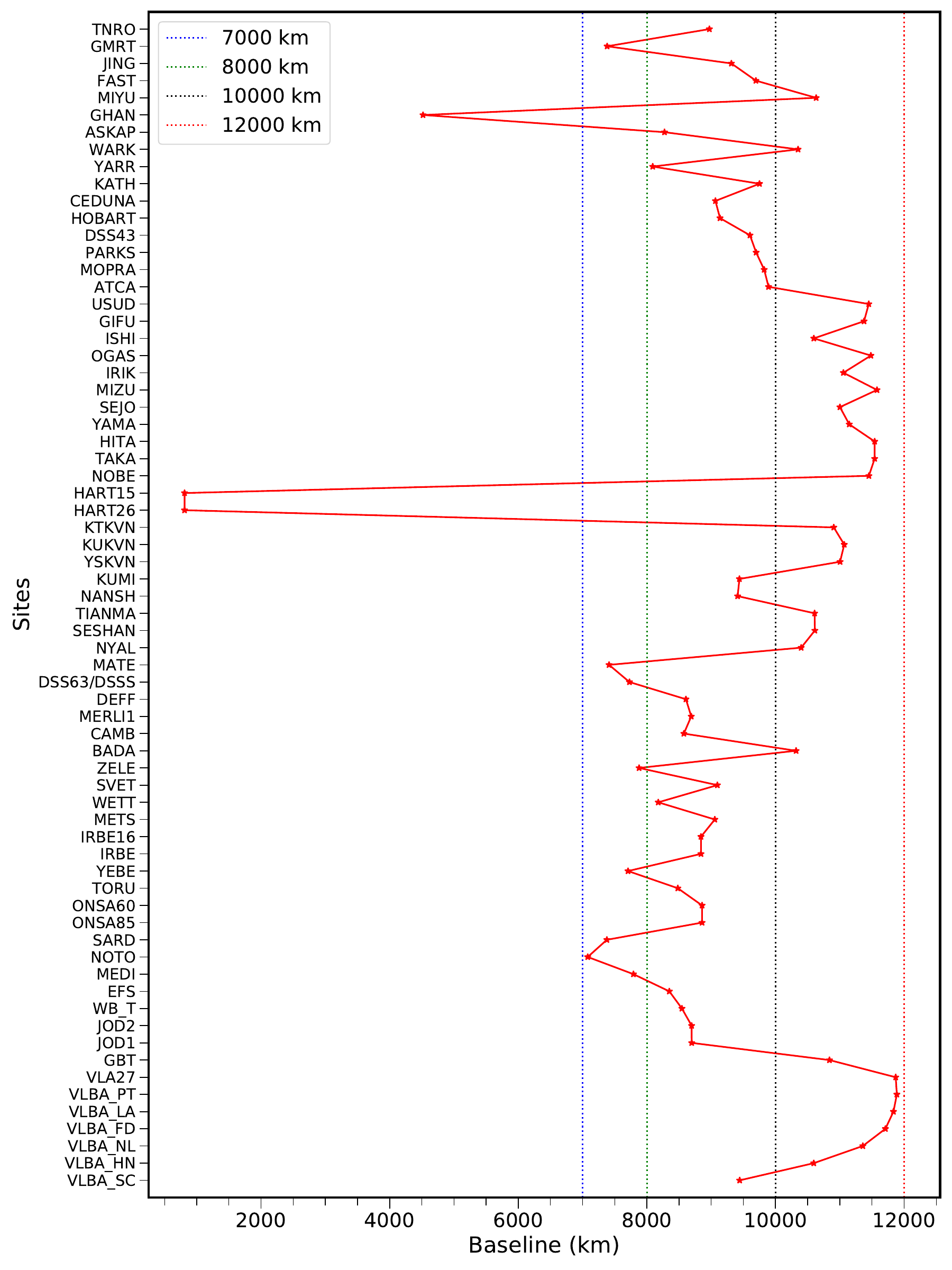}
	\caption{Distribution of the baseline lengths of all sites against SKA-Mid. The blue, green, black and red vertical lines show baselines of 7000, 8000, 10,000 and 12,000 km, respectively.}
	\label{fig:baseline-dist}
\end{figure}

SKA-Mid is suitable for observing sources with Decl. (J2000) below $\sim$$35^{\circ}$, assuming source elevations greater than 20$^{\circ}$. In this work, the scenarios we analyze include sources with Decl. of $-50^{\circ}$, $-35^{\circ}$, $-20^{\circ}$, $-5^{\circ}$, $10^{\circ}$, $25^{\circ}$, and $35^{\circ}$. Considering that not all telescopes that can observe in a certain band can simultaneously observe a source with a specific Decl. (J2000), we divided them into different groups based on UTC time. We showed at most two groups in this work, and more groups can be drawn from Figure \ref{fig:conf} and the distribution of elevation of each telescope (see an example in Figure \ref{fig:elevation}). Figure \ref{fig:uv8} show examples of uv coverage in band C1 for two groups with Decl. of target source being 10$^{\circ}$, where the sites involved in each case are also listed. Table \ref{tab:baseline} lists the corresponding shortest and longest baseline lengths between each site and SKA-Mid. A subclass of facilities able to measure OH (labeled as L\_OH) is also included, because OH masers are important tracers in astronomy. For example, OH masers at 1612/1665/1667 MHz trace diverse physical processes such as star-forming regions and evolved stars \citep[e.g.,][]{Fish-Reid2007, Beuther+2019}, and magnetic fields \citep[e.g.,][]{Fish+2005, Fish-Reid2006, Crutcher+2010}. The 6.7 GHz methanol maser is an important tracer of HMSFRs \citep[e.g.,][]{Ellingsen2006, Green+2009, Yang+2019} and the structure of the MW \citep[e.g.,][]{Xu+2016, Reid+2019, LiJJ+2022}. Figure \ref{fig:baseline-dist} plots the distribution of the baseline lengths of all sites against the SKA-Mid. Most baselines exceed $\sim$8000 km, the maximum length is $\sim$12,000 km, and the typical value is $\sim$10,000 km.

\begin{table}
	\centering
	\scriptsize
	\setlength\tabcolsep{1.5pt}
	\begin{threeparttable}
	\caption{Shortest and Longest Baseline Lengths vs. SKA-Mid\label{tab:baseline}}
	\begin{tabular}{ccccccccccc}
		\hline \hline
		\multicolumn{1}{l}{Decl. } &  L1   &   L2  &  L\_OH   &  C1   &  C2   &  C3   &  X    &  Ku1  &  Ku2  &  Comment \\
		\multicolumn{1}{l}{($^{\circ}$) } &  (km)  &  (km)  &  (km)  &  (km)  &  (km)  &  (km)  &  (km)  &  (km)  &  (km)  &   \\
		\hline
		\multicolumn{11}{c}{Option 1} \\
		\hline
		\multicolumn{1}{l}{$-$50} & (7382, 9892) & (7382, 9892) & (811, 9892) & (811, 10351) & (811, 10351) & (811, 10351) & (811, 10351) & (811, 10351) & ...   & a \\
		\multicolumn{1}{l}{$-$35} & (7382, 10608) & (7382, 10612) & (811, 10612) & (811, 10612) & (811, 10612) & (811, 10612) & (811, 10612) & (811, 10608) & ...   & b \\
		\multicolumn{1}{l}{$-$20} & (7382, 10608) & (7382, 10612) & (811, 10612) & (811, 10612) & (811, 10612) & (811, 11542) & (811, 10612) & (811, 10608) & (9443, 11709) & c \\
		\multicolumn{1}{l}{$-$5} & (7382, 10608) & (7382, 10612) & (811, 10612) & (811, 10612) & (811, 10612) & (811, 11578) & (811, 10632) & (811, 10608) & (8349, 11709) & d \\
		10    & (7080, 10608) & (7080, 10612) & (811, 10612) & (811, 10612) & (811, 10612) & (811, 11578) & (811, 10632) & (811, 10608) & (8349, 11887) & e \\
		25    & (7080, 10608) & (7080, 10612) & (811, 10612) & (811, 10612) & (811, 10612) & (811, 11578) & (811, 10632) & (811, 11358) & (8349, 11358) & f \\
		35    & (7080, 10608) & (7080, 10612) & (811, 10612) & (811, 10612) & (811, 10612) & (811, 11147) & (811, 10632) & (811, 11358) & (8349, 11358) & f \\
		\hline
		\multicolumn{11}{c}{Option 2} \\
		\hline
		\multicolumn{1}{l}{$-$50} & ...   & ...   & ...   & ...   & ...   & ...   & ...   & ...   & ...   & ... \\
		\multicolumn{1}{l}{$-$35} & ...   & ...   & ...   & (811, 9443) & ...   & (811, 9443) & ...   & ...   & ...   & g \\
		\multicolumn{1}{l}{$-$20} & (7080, 7879) & (7080, 7879) & (811, 7879) & (811, 7879) & (811, 7791) & (811, 7791) & (811, 8177) & (9443, 11709) & ...   & h \\
		\multicolumn{1}{l}{$-$5} & (7080, 9411) & (7080, 9411) & (811, 9411) & (811, 9439) & (811, 9439) & (811, 9439) & (811, 9439) & (9443, 11709) & (9443, 11869) & i \\
		10    & (7080, 10839) & (7080, 10839) & (811, 10839) & (811, 10839) & (811, 10839) & (811, 10839) & (811, 10839) & (8349, 11887) & ...   & j \\
		25    & (7080, 10839) & (7080, 10839) & (811, 10839) & (811, 10839) & (811, 11358) & (811, 11358) & (811, 10839) & ...   & ...   & j \\
		35    & (7080, 11358) & (7080, 11358) & (811, 11358) & (811, 11358) & (811, 11358) & (811, 11358) & (811, 11358) & ...   & ...   & j \\
		\hline
	\end{tabular}
	\begin{tablenotes}
		\item{Note.} L\_OH is added to denote the bands that can observe 1612/1665/1667 MHz OH masers. \\
		a. Mainly in Africa, and Oceania. \\
		b. In Africa, Oceania, and Asia. \\
		c. In Africa, Oceania, and Asia in bands L1--Ku1, and in Africa and North America in band Ku2.\\
		d. In Africa, Oceania, and Asia in bands L1--Ku1, and mainly in Africa and North America in band Ku2. \\
		e. Mainly in Africa, Oceania, and Asia in bands L1--X; in Africa, Oceania, Asia, and Europe in band Ku1; and mainly in Africa and North America in band Ku2. \\
		f. Mainly in Africa, Europe, and Asia in bands L1--X, and mainly in Africa and North America in bands Ku1 and Ku2.\\
		g. In Africa and North America (only four sites). \\
		h. Mainly in Africa and Europe in bands L1--X, and in Africa and North America in band Ku1.\\
		i. Mainly in Africa, Europe, and North America in bands L1--X, and in Africa and North America in bands Ku1 and Ku2.\\
		j. Mainly in  Africa, Europe, and North America.
	\end{tablenotes}
\end{threeparttable}
\end{table}

\subsection{Sensitivity and Astrometric Precision Possible with SKA-VLBI} \label{sec:array-sensitivity}

For each pair of antennas, their combined SEFD$_{ij}$ is
\begin{equation}\label{equ:SEFD-baseline}
	\mathrm{SEFD}_{ij} = \sqrt{\mathrm{SEFD}_{i} \times \mathrm{SEFD}_{j}},
\end{equation}
where subscripts $i$ and $j$ present the $i$th and $j$th telescopes, respectively. The total SEFD of an array can read
\begin{equation}\label{equ:SEFD}
	\mathrm{SEFD} = \frac{1}{\sqrt{\sum_{i, j =1}^{n; i < j} \frac{1}{\mathrm{SEFD}_{i}\times \mathrm{SEFD}_{j}}}},
\end{equation}
and the sensitivity for single polarization without considering the effect of sampling is
\begin{equation}\label{equ:sensitivity}
	\Delta S = \frac{\mathrm{SEFD}}{\eta_s} \sqrt{\frac{1}{2 \times \Delta \nu \times \tau_{ff}}},
\end{equation}
where $\Delta \nu$ is the bandwidth, which is set to 128 MHz for the L band (including L1 and L2) and 256 MHz for the other bands \citep[i.e., the same as in][]{Rioja-Dodson2020}; $\tau_{ff}$ is the integration time, which is set to 1 hr; and $\eta_s$ is set to 0.7, i.e., the same as in ``EVN Calculator".\footnote{See \url{https://services.jive.eu/evn-calculator/cgi-bin/EVNcalc.pl}.}

In order to calculate the sensitivity under various conditions, Table \ref{tab:SEFD} lists the SEFDs corresponding to the cases in Table \ref{tab:baseline}. Considering that only the inner 4 km core may be phased up, we expected SEFD of SKA-MID as $\sim$2.6 Jy at $\sim$1--8 GHz \citep[][see more details in Table \ref{tab:telescopes}]{Paragi+2015}. With respect to L2, telescopes that can observe L\_OH add MEER (the upper frequency limit of its L band is 1670 MHz, and the lower limit of its S band is 1750 MHz) and HART26 (the lower frequency limit of its L band is 1608 MHz). The mean SEFDs listed in Table \ref{tab:SEFD}, i.e., the row labeled ``mean,'' were adopted to calculate the corresponding sensitivity of each band (i.e., L1, L2, C1, C2, C3, X, Ku1, and Ku2 at 1.3, 1.6, 5.0, 6.0, 6.7, 8.0, 12.0, and 15.0 GHz, respectively), using Equation (\ref{equ:sensitivity}).

\begin{table}[htbp]
	\centering
	\setlength\tabcolsep{9pt}
	\renewcommand{\arraystretch}{1.3}
	\begin{threeparttable}
		\caption{SEFDs of Possible SKA-VLBI Arrays\label{tab:SEFD}}
		\begin{tabular}{cccccccccc}
			\hline \hline
Decl. & L1 &  L2 & L\_OH  & C1 & C2 & C3 & X & Ku1 & Ku2 \\
($^{\circ}$) & (Jy) & (Jy) & (Jy) & (Jy) & (Jy) & (Jy) & (Jy) & (Jy) & (Jy) \\
			\hline
	\multicolumn{10}{l}{Option 1} \\
\hline
$-$50	&	2.8 	&	3.3 	&	2.5 	&	8.4 	&	8.9 	&	10.3 	&	4.9 	&	17.7	&	...	\\
$-$35	&	1.9 	&	2.1 	&	1.8 	&	3.0 	&	3.1 	&	3.3 	&	4.3 	&	11.2 	&	...	\\
$-$20	&	1.9 	&	2.1 	&	1.8 	&	3.0 	&	3.1 	&	3.1 	&	4.2 	&	11.2 	&	8.3 	\\
$-$5	&	1.2 	&	2.2 	&	1.8 	&	3.0 	&	3.1 	&	3.2 	&	4.4 	&	11.8 	&	7.2 	\\
10   	&	1.1 	&	2.1 	&	1.7 	&	2.6 	&	2.7 	&	2.7 	&	4.4 	&	10.6 	&	4.6 	\\
25	   &	1.0 	&	1.5 	&	1.4 	&	2.0 	&	2.0 	&	2.0 	&	3.3 	&	7.3 	&	7.4 	\\
35     &	1.5 	&	1.5 	&	1.4 	&	2.0 	&	2.0 	&	2.1 	&	3.3 	&	7.3 	&	7.4 	\\
\hline
\multicolumn{10}{l}{Option 2}	\\
\hline	
$-$50	&	...	&	...	&	...	&	... 	&	...	&	... 	&	...	&	...	&	...	\\
$-$35	&	...	&	...	&	...	&	17.9 	&	...	&	19.5 	&	...	&	...	&	...	\\
$-$20	&	3.1 	&	4.5 	&	3.1 	&	10.7	&	9.3 	&	9.0 	&	5.2 	&	8.3 	&	...	\\
$-$5	&	1.5 	&	1.6 	&	1.4 	&	2.1 &	2.2 	&	2.2 	&	3.7 	&	7.2 	&	4.9 	\\
10	    &	1.9 	&	2.0 	&	1.7 	&	2.3 &	2.4 	&	2.4 	&	3.0 	&	4.6 	&	...	\\
25  	&	1.6 	&	1.7 	&	1.5 	&	2.3 &	2.4 	&	2.4 	&	3.0 	&	...	&	...	\\
35	    &	1.6 	&	1.7 	&	1.5 	&	2.3	&	2.4 	&	2.4 	&	2.9 	&	...	&	...	\\
\hline
mean$^a$&	1.8 	&	2.2 	&	1.8 	&	3.6 &	3.6 	&	3.8 	&	3.9 	&	9.7 	&	6.6 \\
			\hline
		\end{tabular}
		\begin{tablenotes}
			\item{Note.} $a$. The case of $-$35$^{\circ}$ in option 2 is not included.
		\end{tablenotes}
	\end{threeparttable}
\end{table}

The position measurement uncertainty (i.e., astrometric precision) reads \citep{Reid+1988, Condon1997}
\begin{equation}\label{equ:astrometry-thermal}
	\Delta \theta \approx 0.5\frac{\theta_{\mathrm{beam}}}{\mathrm{DR}} \approx 0.5 \frac{\lambda}{b} \frac{1}{\mathrm{DR}},
\end{equation}
where $\theta_{\mathrm{beam}} \approx \lambda/b$ is the FWHM of the beam, $b$ is the baseline length, $\lambda$ is the wavelength. Here, $b$ is set to 10,000 km, because the baseline lengths linked to SKA-Mid are mostly around this length (Table \ref{tab:baseline}). Table \ref{tab:astrometry} lists the astrometric precision for DR of 100:1. Note that parallax precision will be 3-5 times better or even better than the astrometric precision with multi-epoch observations \citep[e.g.,][]{Xu+2006, Xu+2013, Deller+2019}. In this work, the parallax precision is assumed to be 1/3 of the astrometric precision.

\begin{figure}[!htb]
	\centering
	\includegraphics[width=1.0\textwidth]{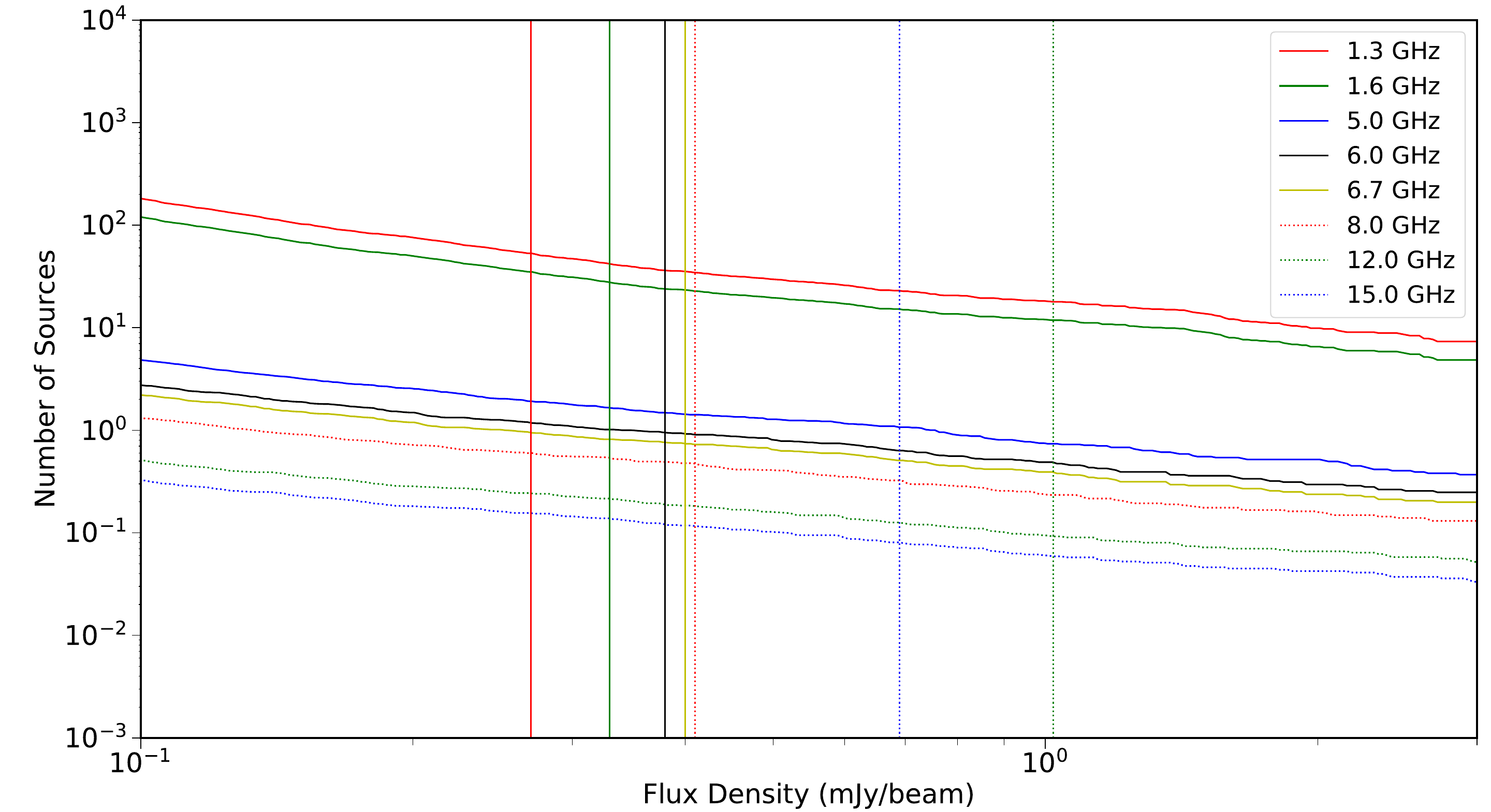}
	\caption{Number of in-beam sources. The vertical lines shows the sensitivity matched DR of 100:1 (see $\Delta S$ values listed in Table \ref{tab:astrometry}), where the lines for 5 and 6 GHz are overlap.}
	\label{fig:source-count}
\end{figure}

We calculated the number of in-beam sources, similar to \citet{Rioja-Dodson2020}. The FoV is \citep{Thompson+2017}
\begin{equation}\label{equ:astrometry-fov}
	\mathrm{FoV} \approx \frac{\lambda}{d},
\end{equation}
where $\lambda$ is the wavelength and $d$ is the aperture, which here is fixed to 32 m because most telescopes have an aperture below 32 m (see Table \ref{tab:telescopes}). Using the tiered radio extragalactic continuum simulation of the continuum radio sky in \citet{Bonaldi+2019}, we can estimate the number of sources that can be detected under different flux densities levels. Their simulations included active galactic nuclei (AGNs) and star-forming galaxies (SFGs), and the number of sources counted in this work includes both objects. In their simulations they only consider frequencies of 1.4, 1.9, 5.0, 6.7, 9.2, 12.5, and 20.0 GHz. Therefore, the calculations in the current work at 1.3 and 1.6 GHz are based on the simulation results at 1.4 GHz, while those at 6.0, 8.0, 12.0, and 15 GHz are based on simulated frequencies of 6.7, 9.2, and 12.5 GHz, respectively (see Figure \ref{fig:source-count}). The numbers of in-beam sources are listed in Table \ref{tab:astrometry}. It is feasible to find multiple in-beam calibrators at L band.

\begin{table}[htbp]
	\centering
	\setlength\tabcolsep{10pt}
	\renewcommand{\arraystretch}{1.3}
	\begin{threeparttable}
		\caption{Characterizing SKA-VLBI Astrometry \label{tab:astrometry}}
		\begin{tabular}{ccccccc}
			\hline \hline
Frequency & SEFD & $\Delta S$  & $\theta_{\mathrm{beam}}$  & $\Delta \theta$ & No. of In-beam  \\
(GHz) & (Jy) & ($\mu$Jy beam$^{-1}$) & (mas) & ($\mu$as) & sources\\
			\hline
1.3 	&  	1.8 	&	2.7 	&	4.8 	&	24	&	53.1	\\
1.6 	& 	2.2 	&	3.3 	&	3.9 	&	19	&	27.7	\\
5.0 	&	3.6 	&	3.8 	&	1.2 	&	6 	&	1.5 	\\
6.0 	&	3.6 	&	3.8 	&	1.0 	&	5	&	1.0 	\\
6.7 	&	3.8 	&	4.0 	&	0.9 	&	5 	&	0.7 	\\
8.0 	&	3.9 	&	4.1 	&	0.8 	&	4	&	0.5 	\\
12.0 	&	9.7 	&	10.2 	&	0.5 	&	3 	&	0.1     \\
15.0 	&	6.6 	&	6.9 	&	0.4 	&	2	&	0.1     \\
			\hline
		\end{tabular}
		\begin{tablenotes}
			\item{Note.} Assuming the baseline length is 10,000 km and the typical telescope aperture is 32 m. For $\Delta S$, the integration time is 1 hr, where the bandwidth is 128 MHz for the L band and 256 MHz for the other bands.
		\end{tablenotes}
	\end{threeparttable}
\end{table}

\section{Astrometric Science with SKA-VLBI}\label{sec:science}

The astrometry of SKA-VLBI will be characterized by high sensitivity and high precision. This section describes some of the astrometric science that can be conducted with SKA-VLBI. The important measurement quantities of astrometry are parallax and proper motion, with which one can obtain 5D (traced by continuum emission) or 6D (combining radial velocities traced by masers) information to study the physical and dynamical properties of target sources. Such objects can be isolated, binary or multiple systems, primary objects plus a group of objects rotating around them, or even more complex systems (such as the MW). In addition, SKA-VLBI can also be used to test GR.

\subsection{Astrometry of Isolated Objects}\label{sec:science-isolate}

SKA-VLBI will provide an opportunity to study isolated stellar objects in different stages of star formation and evolution, e.g., star formation regions (SFRs), post-main-sequence stars (mainly AGB stars), and pulsars \citep[PSRs; e.g., the PSR$\pi$ and MSPSR$\pi$ campaigns;][]{Deller+2011, Vigeland+2018}. Because SFR studies focus on the formation and evolution of central stars, such systems are classified as isolated celestial system rather than multiple systems. Masers, which have been found in SFRs and associated with AGB stars, serve as ideal targets for measuring parallaxes and proper motions, the key quantities of astrometry. For an example online database of masers, please see \citet{Ladeyschikov+2019}.\footnote{\url{https://maserdb.net/}.} The current bottleneck of maser measurements using VLBI, such as with the VLBA, is limitations imposed by limited sensitivity.

\subsubsection{Star Formation Regions}\label{sec:science-isolate-SFR}

\choh~masers at 6.7 GHz were discovered by \citet{Menten1991}, which, along with 22.2 GHz \ho~masers, are the brightest and most widespread maser species in HMSFRs \citep[e.g.,][]{Caswell+1995, Minier+2003, Xu+2008}.
\citet{Sugiyama+2013} conducted 6.7 GHz \choh~maser monitoring studies, using the EAVN to observe 36 target sources in HMSFRs. \citet{Fujisawa+2014} subsequently released the results of this program, who found that maser emission was detected for 35 sources. This group published a detailed case study on the source G006.79-00.25, and found that 6.7 GHz \choh~maser spots were both rotating and expanding\citep{Sugiyama+2016}. It is worth noting that \citet{Moscadelli+2013} pointed out that 6.7 GHz maser emission traces gas entrained in the jet. In addition, \citet{Bartkiewicz+2020} used the EVN to observe 6.7 GHz \choh~masers near the massive YSO G23.657-00.127, and concluded that it was hard to determine whether the methanol masers were tracing a spherical outflow arising from an (almost) edge on disk or a wide-angle wind at the base of a protostellar jet. SKA-VLBI will have high sensitivity and will be able to detect faint sources. Assuming an observation at 6.7 GHz with an integration time of 1 hr and a bandwidth of 2 kHz, a 1$\sigma$ would be possible at $\sim$1.4 mJy beam$^{-1}$ (see Table \ref{tab:astrometry}), which is better than the $\sim$4 mJy beam$^{-1}$ in \citet{Bartkiewicz+2020}. The estimated beam size ($\sim$0.9 mas) is also smaller than that in \citet{Bartkiewicz+2020} by a factor of $\sim$7. These facts indicate it is very likely that more maser spots will be detected by SKA-VLBI, and the resulting astrometric precision will also improve. Combining search of radio continuum emission with high spatial resolution (such as by SKA-Mid) to pinpoint the driving source of the masers, SKA-VLBI is expected to discern between the scenarios of sphere-like outflow and wide angle wind at the base of the protostellar jet \citep[see above,][]{Bartkiewicz+2020}, i.e., distinguish which part of the outflowing gas near YSO is traced by 6.7 GHz \choh~masers. It is also possible to conduct further similar studies on more HMSFRs to determine statistically whether the scenario is unified for different sources.

Gould's Belt Distances Survey (GOBELINS) was dedicated to measuring the distances and proper motions of young stars in neighboring molecular clouds using VLBA at 5 and/or 8 GHz with a bandwidth of 256 MHz. Results from this survey have been reported for the Ophiuchus complex \citep{Ortiz-Leon+2017}, Orion molecular clouds \citep{Kounkel+2017}, the Serpens/Aquila molecular complex \citep{Ortiz-Leon+2017b}, and the Perseus molecular cloud \citep{Ortiz-Leon+2018}. The dynamics of star formation and young stars within molecular clouds were studied. For binary or multiple young stars, physical properties such as mass and orbital parameters have also been published (for more details see Section \ref{sec:science-binary}). The typical sensitivity of GOBELINS was $
\sim$25 $\mu$Jy, which is far worse than the $\Delta S$ in Table \ref{tab:astrometry}, indicating that SKA-VLBI will be able to detect more younger stars \citep[see also][]{Ortiz-Leon+2017}. Therefore, SKA-VLBI is expected to help compose a more complete picture of the synergistic study of young stars and molecular cloud dynamics.

In addition, it is interesting to note that in a case study of the Protostellar Outflows at the EarliesT Stages (POETS) survey presented by \citet{Sanna+2019}, linearly polarized emission from 6.7 GHz \choh~masers was detected for G035.02+0.35, and the direction of the magnetic field was found to be consistent with the axis of the jet. It is expected that studies of the Zeeman effect in 6.7 GHz \choh~masers \citep[e.g.,][]{Vlemmings+2008, Crutcher+2019, Dall'Olio+2020} of early stage protostars will further complement studies of their magnetic fields by using SKA-VLBI with its higher sensitivity. This will enable the study of the magnetodynamics of SFRs and the influence of magnetic fields on star formation.

\subsubsection{Post-Main-Sequence Stars}\label{sec:science-isolate-PMSS}

The significance of multiband and multiepoch observations of post-main-sequence stars (e.g., AGB stars, red supergiant stars, etc.) is that they can be used to (i) investigate the astrochemistry of material that has been ejected into the interstellar medium (ISM); (ii) distinguish different maser pumping mechanisms (e.g., radiative or collisional pumping), and derive these physical properties accordingly in circumstellar envelopes (CSEs); (iii) study the late stages of stellar evolution; and (iv) probe the spatial structure and dynamics related to mass-loss processes and measure asymmetry in CSEs \citep[see][]{Reid-Honma2014, Rioja-Dodson2020}. The study of these post-main-sequence stars mainly relies on observations of 22.2 GHz \ho, 43.1, 42.8, 86.2, and 129.3 GHz SiO, 1612, 1665, and 1667 MHz OH masers, and others, which are used to measure parallaxes, proper motions, radial velocities, and light variations, and to determine the overlaps or offsets between masers of different transitions, ring sizes, and their temporal evolution \citep{Reid-Honma2014}. The VLBI parallaxes of AGB stars can also be compared with Gaia parallaxes, which are used to constrain the Gaia parallax zero-point offset (see Section \ref{sec:science-isolate-RS} for details), and to calibrate the period-luminosity relation derived from oxygen-rich Mira variables in the MW \citep{Andriantsaralaza+2022}.

Below 15 GHz, OH masers are good tracers of AGB stars. OH masers at 1612 MHz are a great tracer of low- and intermediate-mass oxygen-rich evolved stars with long-period stellar pulsations (i.e., types of AGB stars). This type of evolutionary stage with strong 1612 MHz OH maser emission is exclusively attributed to OH/IR stars. Observations of OH/IR stars can be traced back to the 1980s \citep[e.g.,][]{Herman-Habing+1985, van-Langevelde+1990, Lindqvist+1992}, and include large-scale sky surveys performed with the Australian Telescope Compact Array (ATCA) and VLA \citep{Sevenster-Chapman+1997, Sevenster+1997, Sevenster+2001}. \citet{Engels+2015} provided a database for querying known 1612, 1665, and 1667 MHz OH masers.\footnote{See \url{https://hsweb.hs.uni-hamburg.de/projects/maserdb/}.} Combining the HI/OH/Recombination line (THOR) survey of the MW \citep{Beuther+2016, Beuther+2019} and follow-up observations by the Southern Parkes Large-Area Survey by the ATCA in Hydroxyl \citep[SPLASH;][]{Qiao+2016, Qiao+2018, Qiao+2020}, \citet{Uno+2021} conducted 3D simulations of OH/IR stars traced by 1612 MHz OH masers in the Galactic disk. From these simulation data, and within a scale height of 150 pc, they found there may $\sim$3990 OH/IR stars in the Galactic plane, for which they determined their kinematics distances.

\begin{figure}[!htb]
	\centering
	\begin{subfigure}{0.495\textwidth}
		\includegraphics[width=1.0\textwidth]{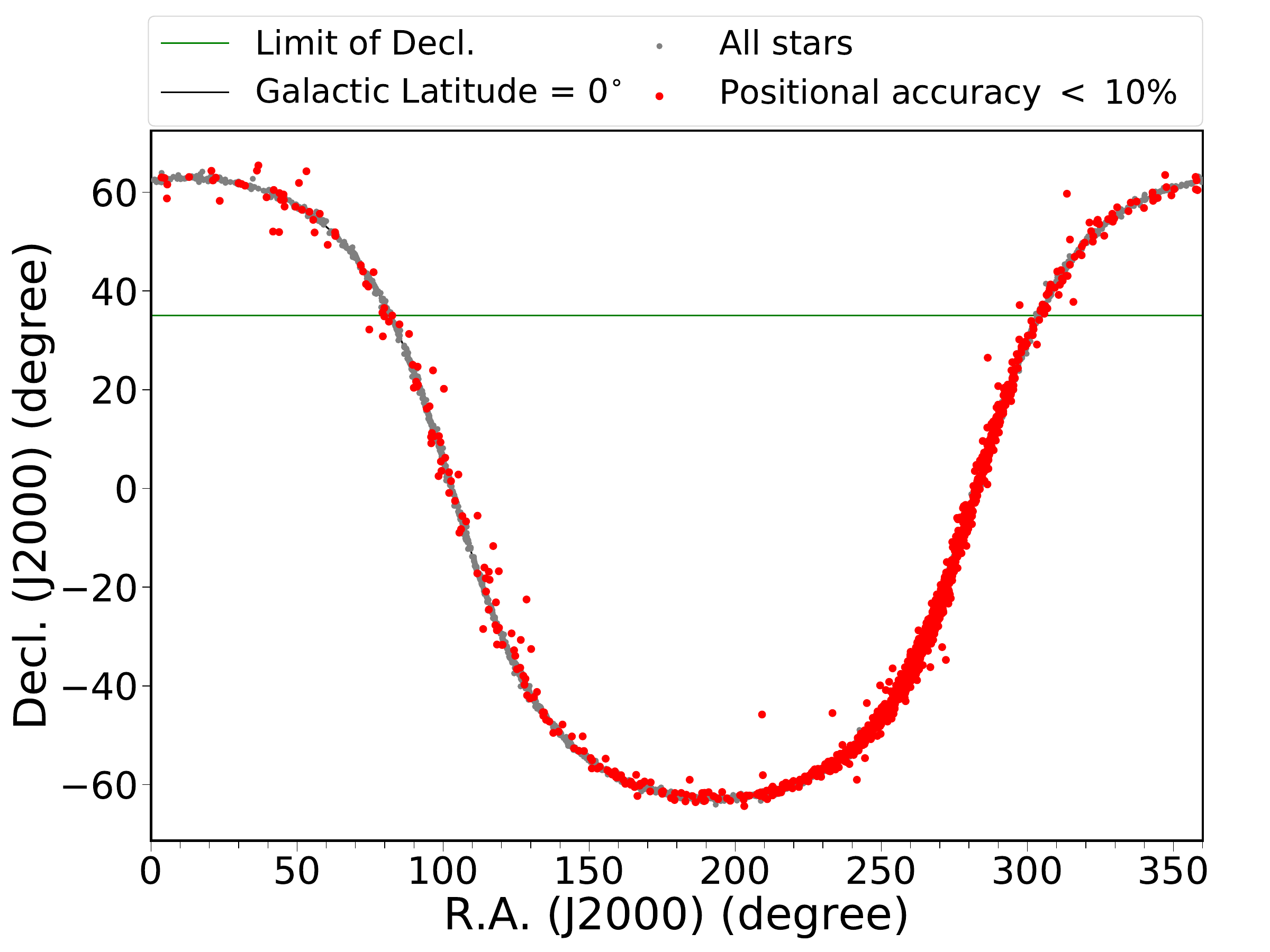}
		\caption{All stars.}
	\end{subfigure}
	\begin{subfigure}{0.495\textwidth}
	     \includegraphics[width=1.0\textwidth]{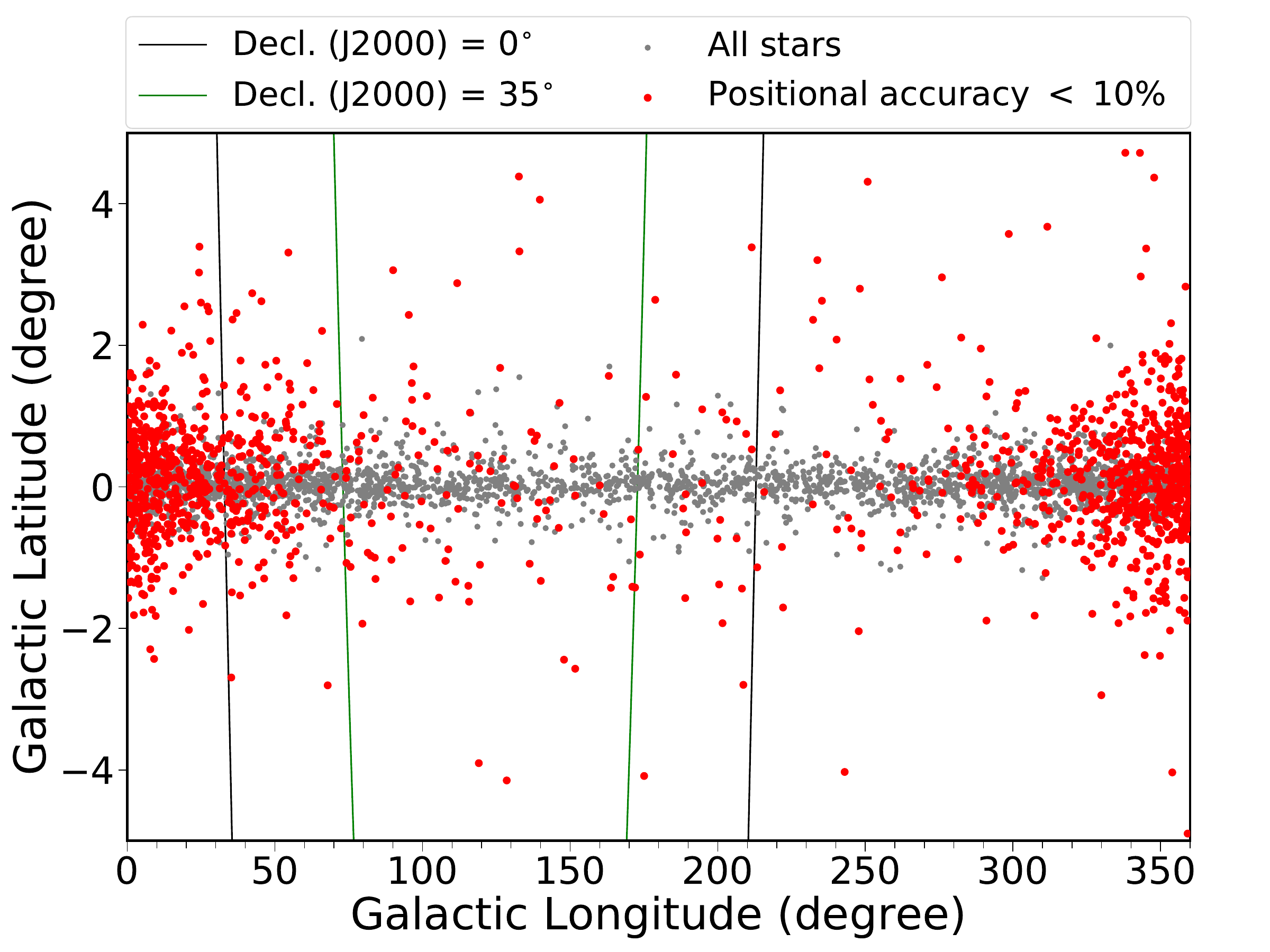}
	     \caption{Zoom-in of stars within a Galactic latitude of $|5^{\circ}|$.}
    \end{subfigure}
	\caption{The distribution of OH/IR stars housing 1612 MHz OH masers from \citet{Uno+2021}, where the gray and red points present all OH/IR stars and stars with a theoretical parallax precision better than 10\%, respectively. In the left panel, the black line presents a Galactic latitude of 0$^{\circ}$, and the green line indicates the Decl. (J2000) limit of SKA-VLBI, i.e., 35$^{\circ}$. In the right panel, the black and green lines present Decl. (J2000) of 0$^{\circ}$ and 35$^{\circ}$, respectively.
	}
	\label{fig:science-post-ms}
\end{figure}

Ionospheric residuals can be largely reduced using MultiView, and a spatial resolution can reach $\sim$3.9 mas, thus rendering SKA-VLBI a suitable instrument for studying OH/IR stars traced by 1612 MHz OH masers. The specific technical methodology and scientific strategy for future SKA astrometric missions to study 1612 MHz OH masers is currently under development \citep[see][]{Orosz+2014}. Figure \ref{fig:science-post-ms} shows the distribution of $\sim$3990 OH/IR stars, and the results show that SKA-VLBI can measure, in theory, the parallaxes of about 34\% (1345/3990) of OH/IR stars with a distance precision of better than 10\% (here, we have assumed a bandwidth of 500 Hz, an integration time of 1 hr, the astrometric properties in Table \ref{tab:astrometry}, the highest parallax precision of 6 $\mu$as, and the maximum Decl. (J2000) of 35$^{\circ}$). OH masers at 1665 and/or 1667 MHz can also be used to study AGB stars \citep[e.g.,][]{van-Langevelde2000, Vlemmings+2003, Vlemmings+2007}.

\subsubsection{Pulsars}\label{sec:science-isolate-PS}

\citet{Deller+2011} proposed a major project to measure the parallaxes of pulsars (some of which are in binary systems), PSR$\pi$, using VLBA. This project planed to observe 60 and 200 pulsars in its Phases I and II, respectively. Phase I was implemented over a 9 yr period \citep{Deller+2016, Deller+2019} and the parallaxes and proper motions of 60 pulsars were obtained. The typical parallax precision was $\sim$45 $\mu$as, with the best being close to 10 $\mu$as, which was much more precise and accurate than pulsar's dispersion measure (DM)-based distance \citep{Chatterjee+2009, Deller+2009, Deller+2019}. It is worth noting that the astrometric errors of these near but weak sources are mainly determined by thermal noise \citep{Rioja-Dodson2020}. \citet{Vigeland+2018} reported a similar project, i.e., observing millisecond pulsars (which are highly probably in binary systems), MSPSR$\pi$, using VLBA, and the parallaxes and proper motions of 18 millisecond pulsars were reported \citep{Ding+2020, Ding+2023}.\footnote{For the specific status of PSR$\pi$ and MSPSR$\pi$, see \url{https://safe.nrao.edu/vlba/psrpi/status.html}.}

In addition to measuring the parallaxes and proper motions of (millisecond) pulsars, a series of studies, including those related to projects PSR$\pi$ and MSPSR$\pi$, have made great progress obtaining the physical properties of pulsars, e.g., their effective temperature, radius, mass, and average electron density, and provided constrains on the time derivative of Newton's gravitational constant \citep[e.g.,][]{Vigeland+2018, Deller+2019, Ding+2020, Ding+2021, Ding+2023}. This series of work has also contributed to the development of pulsar timing array (PTA) studies, such as providing accurate distances to pulsars, improving PTA sensitivity, and verifying and improving pulsar timing solutions \citep{Madison+2013, Vigeland+2018, Deller+2019}. At the same time, pulsars provide a rich laboratory for studying nuclear and particle physics \citep{Lorimer+2008, Deller+2011}; they can be used to test GR \citep{Deller+2009, Deller+2018, Ding+2021, Guo+2021}; they were used to find the first observational evidence for the existence of gravitational waves \citep{Taylor-Weisberg1989}; and their study led to the discovery of the first exoplanet \citep{Wolszczan-Frail1992}.

\begin{figure}
	\centering
	\includegraphics[width=1.0\textwidth]{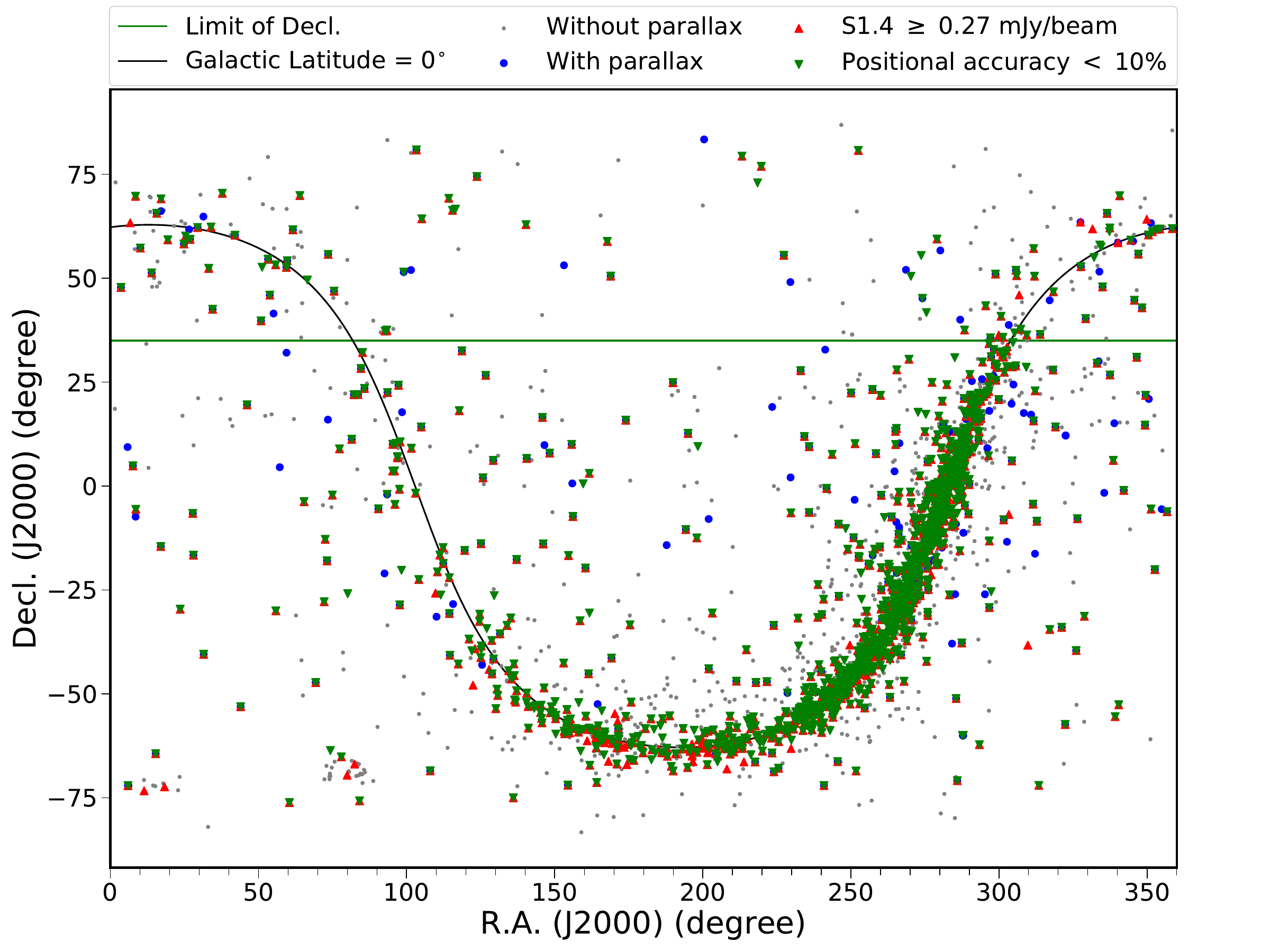}
	\caption{The distribution of pulsars from \citet{Manchester+2016}, where the gray and blue points present pulsars with and without a parallax measurement, respectively. The red upper and green lower triangles indicate objects with flux densities at 1.4 GHz $\geq$ 0.27 mJy beam$^{-1}$ (i.e., the sensitivity matched to the DR of 100:1, see $\Delta S$ in Table \ref{tab:astrometry}) and objects with a theoretical parallax precision better than 10\%, respectively. The black line presents a Galactic latitude of 0$^{\circ}$, and the green line indicates the Decl. (J2000) limit, 35$^{\circ}$, of SKA-VLBI.}
	\label{fig:science-psr}
\end{figure}

The advantages of MultiView are required for determining precise astrometry for pulsars, so that stronger sources with larger angular distances can be used as reference calibrators, providing commensurate precision \citep[see][]{Rioja-Dodson2020}. The high sensitivity of SKA-VLBI not only will enable the detection of weak sources while maintaining high DR, but its will also enable in-beam MultiView observations. In this section, the pulsar astrometry presented here is based on the assumption that astrometric precision is dominated by thermal noise. Herein, we consider the pulsar catalog provided by \citet{Manchester+2005} and recently updated by \citep{Manchester+2016}, which contains 2536 pulsars. (Note that the pulsars published in PSR$\pi$ and MSPSR$\pi$ are also in this catalog.) Figure \ref{fig:science-psr} shows the distribution of these pulsars, of which 12\% (295/2536) have parallax and/or proper motion measurements. In the catalog of \citet{Manchester+2016}, 1671 pulsars have flux density measurements (at 1.4 GHz), and 2481 have distance measurements (default as DM-based distance), but 864 have no flux density or distance measurements. According to the information in the catalog and the astrometric properties listed in Table \ref{tab:astrometry}, in theory, a total of $\sim$48\% (1225/2536) of these pulsars, located with a Decl. (J2000) below 35$^{\circ}$, can be measured with a distance precision better than 10\%, and sources can be detected by to 12.5 kpc (where the highest parallax precision is assumed to be $\sim$8 $\mu$as, i.e., 1/3 of the astrometry precision listed in Table \ref{tab:astrometry}). There are 1014 pulsars whose flux densities are higher than the sensitivity matched to the DRs of 100:1 (i.e., $\geq$ 0.27 mJy beam$^{-1}$) and Decl. below 35$^{\circ}$. Note that the observed DRs of pulsars can be enhanced when pulsar gating is applied in the correlator \citep[see][]{Deller+2019, Ding+2023}. It is undeniable that SKA-VLBI astrometry with high sensitivity and high precision will be able to observe a large number of pulsars, making contributions to the measurement of electron density in the MW and the other science goals mentioned above.

\subsubsection{Radio Stars: VLBI versus Gaia Parallaxes}\label{sec:science-isolate-RS}

Radio stars are stars visible in the radio band, which contain young stars, AGB stars, pulsars, etc. Multiepoch SKA-VLBI observations of radio stars can yield both high-precision astrometric parameters (such as parallax and proper motion) and high spatial resolution images. Combined with optical Gaia astrometry, the high-precision astrometry of SKA-VLBI can be used to link the optical and radio celestial reference frames, and verify the astrometric performance of VLBI and Gaia.

Long-term astronomical observations in both radio and optical bands can produce celestial reference frames, such as ICRF3 \citep{Charlot2020} for radio and the Gaia DR2/(E)DR3 celestial reference frame \citep[Gaia-CRF2/CRF3, which is collectively called GCRF;][]{Gaia-Collaboration+2018, Gaia-Collaboration+2022} in the optical. As mentioned previously, there are also notable VLBI-Gaia positional offsets \citep{Mignard+2016, Kovalev+2017, Petrov+2017, Charlot2020, Liu-Lambert+2021, Secrest+2022}. Observations of radio stars are helpful to tie together these international celestial reference frames \citep[][for more details see Section \ref{sec:science-complex-ICRF}]{Reid-Honma2014, Malkin2016}.

It is interesting to compare the parallaxes of evolved AGB stars and other radio stars obtained with VLBI \citep[i.e., such as the more than 130 stars used in the study by][]{Xu-Zhang+2019} with those obtained with Gaia, as such as comparison offers a unique opportunity to test the Gaia parallax zero-point offset. \citet{Xu-Zhang+2019} compared the VLBI and Gaia parallaxes of their sample of stars and found that except for AGB stars, there was a systematic offset between the parallaxes, $-75 \pm 29$ $\mu$as, which was consistent with the zero-point offset of Gaia. For AGB stars, there was a significant difference between the Gaia and VLBI parallaxes. \citet{Andriantsaralaza+2022} compared the VLBI parallaxes of AGB stars with those of Gaia DR3 and obtained a zero-point offset of DR3 parallax of $\sim -77$ $\mu$as for bright AGB stars. They also found that the zero-point offset become more negative for fainter AGB stars, and Gaia DR3 underestimated the uncertainties of the parallaxes of these AGB stars.

\begin{figure}[!htb]
	\centering
	\includegraphics[width=1.0\textwidth]{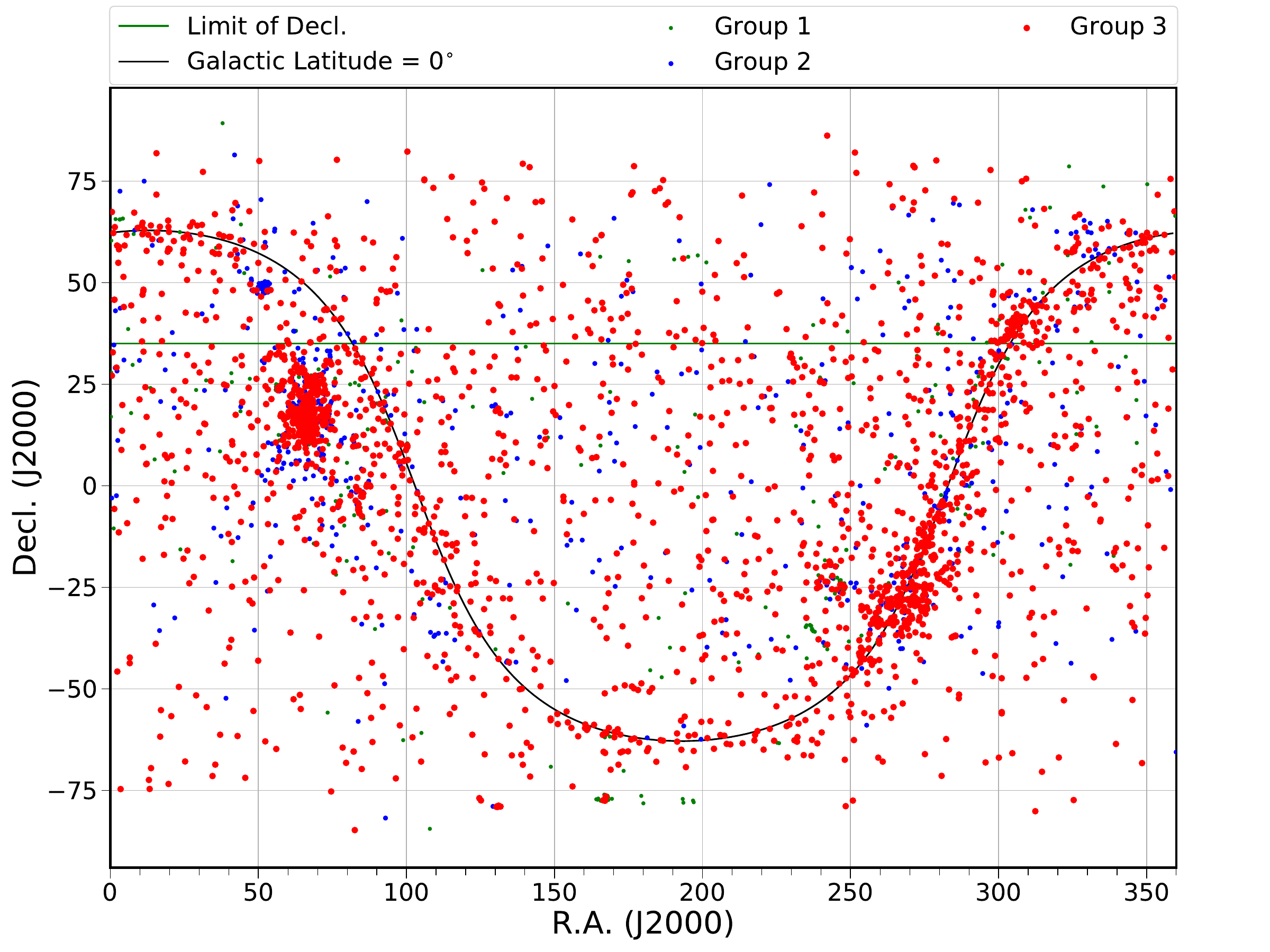}
	\caption{The distribution of radio stars from \citet{Wendker+1995}, where the green points indicate all radio stars (Group 1), the blue points present sources that had been observed at the frequencies considered in this work (Group 2), and the red points show sources whose flux densities (or upper limits) are higher than the sensitivity matched to the DR of 100:1 (Group3). The black line presents a Galactic latitude of 0$^{\circ}$, and the green line indicates the limiting Decl. (J2000) of SKA-VLBI, i.e., 35$^{\circ}$.}
		\label{fig:science-rs}
\end{figure}

Radio star astrometric observation plans were made in the latter part of the last century \citep[e.g.,][]{Lestrade+1986, Garrington+1995, Lestrade+1999, Johnston+2003} and the early part of this century \citep[e.g.,][]{Boboltz+2003, Boboltz+2007}. \citet{Wendker+1995} provided a catalog of radio stars (note that the sample of radio stars discussed in this passage specifically refers to stars from this catalog) which is continually updated and currently contains 3699 stars (their distribution is displayed in Figure \ref{fig:science-rs}). These radio stars were observed at frequencies ranging from 20 MHz to 420 GHz, with 3230 stars falling within the SKA-VLBI frequency range (which will be given priority in this work).
SKA-VLBI will be able to measure precisely the parallaxes of radio stars whose flux densities (or upper limits) are higher than the sensitivity matched to the DR of 100:1 (only considering sources observed at 1.0--2.0 and 4.6--15.3 GHz and had a Decl.(J2000) $<$ 35$^{\circ}$). The number of such radio stars is $\sim$2033, which is about one order of magnitude higher than the number of radio stars that have reported VLBI parallaxes.
	
\subsubsection{Other Isolated Objects}\label{sec:science-isolate-others}

\citet{Matthews+2023} reported the detection of 15 GHz continuous emission for the first time using VLA observations of a classic Cepheid, $\delta$ Cephei ($\delta$ Cep; distance of $\sim$0.3 kpc), with a flux density $\sim$15 $\mu$Jy. For an integration time of 4 hr and a bandwidth of 1 GHz, the corresponding sensitivity of SKA-VLBI is $\sim$1.8 $\mu$Jy beam$^{-1}$. SKA-VLBI has the potential for direct triangular parallax measurements of classical Cepheids to be conducted, allowing astronomers to calibrate the period-luminosity relation more precisely. The theoretical distance precision is better than 10\% for a classic Cepheid within 1.0 kpc assuming that its flux density at 0.3 kpc is $\sim$15 $\mu$Jy at 15 GHz (as in the case of $\delta$ Cep), where the assumed flux density is inversely proportional to the square of distance. \citet{Pietrukowicz+2021} cataloged classical Cepheids in the MW, updated as of 2023,\footnote{See \url{https://www.astrouw.edu.pl/ogle/ogle4/OCVS/allGalCep.listID}.} and these classical Cepheids have been cross-matched with Gaia DR3 \citep{Gaia-Collaboration-DR3-2022}. There are over 3000 classical Cepheids in this catalog, with $\sim$80 within $\sim$1.0 kpc. \citet{Ripepi+2022} provided a Gaia DR3 catalog of Cepheids containing some in neighboring galaxies, totaling over 15,000 Cepheids.

Combining high-precision VLBI astrometry and lunar ranging data, the mass of the Earth-Moon system and the Moon's momentum of inertia ratio can be measured \citep[see][]{Lanyi+2007, Reid-Honma2014, Reid+2019baas}. Combining VLBI data with optical observations, Doppler ranging data, and data from the Cassini satellite, the mass and potential of the Saturnian system were constrained  \citep[e.g.,][]{Jacobson+2006}. In addition, high-precision VLBI astrometry can also be used to monitor the trajectories of artificial radio sources (such as man-made spacecraft), contribute to geophysical research, \citep[such as measuring high-altitude winds, see][]{Counselman+1979, Sagdeyev+1992, Reid-Honma2014}, and improve future space missions \citep[see][]{Duev+2012}. More research on satellite tracking can be found in the review by \citet{Reid-Honma2014}.

\subsection{Astrometry of Binary and Multiple  Systems}\label{sec:science-binary-Multi}

\subsubsection{Binary or Multiple Stellar Systems}\label{sec:science-binary}

Binary or multiple stellar systems include young binary or multiple stars, X-ray binaries (where one of the stars is a stellar black hole, a neutron star, or a white dwarf), Hulse-Taylor binary pulsar systems \citep[HTBPSRs, which are composed of a neutron star and a pulsar, see][]{Hulse-Taylor1975}, and other binary or multiple stellar systems containing neutron stars or black holes, cataclysmic variables (CVs, where the primary star is a white dwarf); see more details below. It is possible to monitor the orbital motions and measure the parallaxes, masses, orbital periods, and other physical quantities of these systems. The significance of studying binary or multiple stars includes testing theories of stellar structure and evolution; exploring star formation mechanisms and environmental effects in the galactic gravitational potential and in clusters; understanding important types of astrophysical phenomena such as Type Ia supernovae, CVs, and stellar X-ray sources \citep{Horch2013}; and testing fundamental physical theories such as the relativistic effects of binary/multiple stellar systems containing neutron stars and/or black holes \citep[e.g.,][]{Deller+2018, Ding+2021, Guo+2021}; etc.

\textit{Younger binary or multiple stars}. Orbital monitoring of young binary stars using VLA/VLBA has been carried out \citep[e.g.,][and GOBELINS, see Section \ref{sec:science-isolate-SFR}]{Curiel+2002, Loinard2002, Rodriguez+2003, Maureira+2020}. These studies provided orbital and other physical parameters of the binary/multiple-star systems. For example, \citet{Rodriguez+2003} obtained the total mass and orbital period of the binary L1551 IRS 5 using the VLA, finding $\sim$1.2 $M_{\odot}$ and $\sim$260 yr, while \citet{Maureira+2020} combined VLA observations from \citet{Loinard2002} and ALMA data to derive the mass range of the binary IRAS 16293-2422 A, finding $0.5 \lesssim M_1 \lesssim M_2 \lesssim 2\; M_{\odot}$. Using VLBA with its higher astrometric precision, GOBELINS also monitored the orbits of 17 young binary and nine young multiple stars, and obtained the orbital parameters for a total of 12 binary stars, including the mass of each single star in the binary system for nine systems. The maximum precision reached $\sim$0.02 $M_{\odot}$ \citep[corresponding to a relative precision up to better than 10\%; see][]{Kounkel+2017, Ortiz-Leon+2017, Ortiz-Leon+2017b, Ortiz-Leon+2018}. Among them, \citet{Ortiz-Leon+2017} even determined the mass of two stars in a triple-star system, YLW 12B.
The value of $\theta_{\mathrm{beam}}$ at 5.0 GHz is $\sim$1.2 mas (see Table \ref{tab:astrometry}), and the astrometric precision is better than 11 $\mu$as \citep[assuming the flux density at 5 GHz is better than the minimum continuous emission of $\sim$0.2 mJy in GOBELINS;][i.e., the matched DR is $\sim$53:1, as shown in Table \ref{tab:astrometry}]{Ortiz-Leon+2017}. Both of these are less than the semimajor axes of binary younger stars reported in GOBELINS (i.e., over $\sim$10 mas). Therefore, the astrometric precision that may be obtained using SKA-VLBI may be sufficient to conduct orbital monitoring of such binary and multiple star systems and more (see Section \ref{sec:science-isolate-SFR}).

\textit{X-ray binaries}. Some X-ray binaries have properties similar to quasars or other AGNs, but as they evolve much faster, they are often called micro-quasars. X-ray binaries can be used to study the interaction of stars with the accretion disks of the black holes or neutron stars they orbit, and to study the stability and lifetime of the corresponding jets; such approaches have been useful for studying the relativistic jets of distant quasars \citep{Mirabel+1998, Reynolds2021}. An important example of VLBI observation of X-ray binary is Cygnus X-1 \citep{Reid+2011, Orosz+2011, Miller-Jones+2021}; after obtaining the distance to this X-ray binary and monitoring its orbit, the mass-distance degeneracy in the simulation of optical/infrared data was broken, and the mass of the black hole and its companion star was therefore obtained.
In addition, the parallaxes of X-ray binaries obtained with VLBI can be compared with the distances obtained from optical observations, which can be used to constrain theories such as accretion disk theory. For example, the distance to SS Cyg measured by EVN and VLBA \citep{Miller-Jones+2013} was smaller than that determined using the Hubble Space Telescope observations \citep{Harrison+1999}, making the distance more consistent with accuracy disk theory \citep[for a review see][]{Reid-Honma2014}. Several examples of X-ray binaries with measured parallaxes include Sco X-1, V404 Cyg, Cygnus X-1, and SS Cyg \citep[see][]{Fomalont+2001, Miller-Jones+2009, Reid+2011, Miller-Jones+2013}, which were observed at frequencies of 1.7--22.2 GHz and their measured flux densities were $\gtrsim$ 5 mJy beam$^{-1}$. It is expected that SKA-VLBI will be able to measure the parallaxes and semimajor axes of primary stars with a theoretical parallax precision of $\sim$1 $\mu$as. It is also expected that SKA-VLBI will enable the exploration of fainter X-ray binaries and allow astronomers to monitor their light curves \citep[e.g.,][]{Ng+2023} with high sensitivity. Samples of X-ray binaries can be obtained from the Galactic high mass X-ray binary catalogue \citep{Neumann+2023} and the Galactic low mass X-ray binary catalogue \citep{Avakyan+2023}.

\textit{HTBPSRs}. An HTBPSR is a specific type of binary containing a pulsar \citep{Hulse-Taylor1975}. Such binaries led to the first indirect detection of gravitational waves \citep[see][]{Taylor-Weisberg1982, Taylor-Weisberg1989} and can also be used to test GR (see Section \ref{sec:science-GR-other} for details). An important representative is pulsar B1913+16, which has been studied extensively for many years. \citet{Weisberg-Taylor2005} made an important review of studies of B1963+16, and since then there have been observations conducted by large radio telescopes and arrays \citep[e.g., Arecibo and VLBA;][]{Nice+2005, Deller+2018} and analyses of long-term observational data \citep[e.g.,][]{Weisberg-Huang2016}. \citet{Weisberg-Huang2016} obtained the ratio of the observed to GR-predicted orbital period decay rate (ROPR hereafter) of B1913+16 due to gravitational wave damping, $0.9983 \pm 0.0016$, which tested the GR with an precision of 0.2\%. They also found that the result was also consistent with the GR-predicted value from the Shapiro gravitational propagation delay. Using eight epochs of observations made over 2 yr with VLBA, \citet{Deller+2018} obtained the ROPR, finding a value of $1.005^{+0.001}_{-0.003}$, which was incongruous with that of \citet{Weisberg-Huang2016} but with similar level of precision. Therefore, further radio VLBI observations with higher astrometric precision than achieved for pulsar B1913+16 and observations toward more HTBPSRs (which may be fainter) are imperative. These, in conjunction with pulsar observations (see Section \ref{sec:science-isolate-PS}), will constitute an important area of focus for SKA-VLBI.

\textit{Cataclysmic variables.} CVs are interacting binaries in which a white dwarf accretes material from a low-mass companion. CVs are valuable to test theories of the evolution of binary stars; however, it is difficult to obtain precise values of the masses and radii of each star in a CV \citep{Knigge+2011, Littlefair+2024}. VLBI detections of CVs are scarce \citep[e.g., nova RS Ophiuchi by VLBA;][see nova below]{O'Brien2006}. Recently, \citet{Jiang-Cui+2023} reported an EVN detection of an AE Aquarii (AE Aqr)-type CV,\footnote{It is a subtype of intermediate polars, which are a subclass of CVs with a moderate magnetic field strength of $10^5 \lesssim B \lesssim 10^7$ G and a sufficiently fast spin rate.} LAMOST J024048.51+195226.9, which possesses the fastest known rotating white dwarf with a spin period of $\sim$25 s. This paper reported only its equatorial coordinates and no other astrometric parameter. Its maximum flux density at 1.7 GHz was $\sim$1.0 mJy, indicating a potential target for SKA-VLBI to explore.

\textit{Novae}. A nova is also a kind of CV. A classical nova is an eruption event on the surface of an accreting white dwarf binary. In such an event, the typical released energy is $\sim$10$^{38}$--10$^{43}$ erg, and heavy elements inside a white dwarf are ejected into the ISM \citep{Gallagher-Starrfield1978, Gehrz+1998, Jose-Hernanz2007, Romano+2017}, implying novae have a significant impact on the ISMs. Currently, theory and observations of novae are not overly consistent, especially in terms of the ejecta masses of different subtypes of novae. Radio observations are an ideal tracer of ejecta masses, and have already made significant contributions to the study of novae, such as constraining the expanding ejecta and their morphologies, distances, shell masses, and kinetic energies \citep[see the short review by][]{Gulati+2023}. Gulati et al. conducted a systematic search for novae with Decl. $\leq$ 41$^{\circ}$ using ASKAP, discovering four nova radio counterparts, and conducted follow-up observations using ATCA. Their results highlight ASKAP's ability to contribute to future radio research of novae. The maximum radio flux density of a nova is related to the ejected mass, $M_{\mathrm{ej}}$. At 1 GHz, assuming $M_{\mathrm{ej}} \sim 5 \times 10^{-4}$ $M_{\odot}$, SKA-VLBI may, in theory, be able to locate sources within $\sim$10 kpc with an precision of better than 10\% (based on the astrometry in Table \ref{tab:astrometry}) and the maximum flux density, which is associated with the $M_{\mathrm{ej}}$ taken from table 4 in \citealt{Gulati+2023}).\footnote{For example, the maximum flux density at 1.0 GHz for a novae at 0.8 kpc is $\sim$ 37.71, $\sim$ 5.98 and $\sim$ 0.95 mJy for $M_{\mathrm{ej}} \sim 5 \times 10^{-4}$, $\sim 5 \times 10^{-5}$ and $\sim 5 \times 10^{-6}$ $M_{\odot}$, respectively.} Note that the assumed maximum flux density is inversely proportional to the square of the distance. If the values of $M_{\mathrm{ej}}$ are $5 \times 10^{-5}$ and $5 \times 10^{-6}$ $M_{\odot}$, the corresponding distance becomes $\sim$6 and $\sim$3 kpc, respectively. In these cases, the corresponding mass of the white dwarf is $\sim$0.4--1.25 $M_{\odot}$ \citep{Yaron+2005, Gulati+2023}.

\textit{Electromagnetic counterparts to gravitational wave sources}. Such sources correspond to merge of two neutron stars (NSNS), or a neutron star and a black hole (NSBH). A correlation linking short gamma-ray bursts, kilonovae, and NSNS/NSBH merges can be found in figure 1 of \citet{Fernandez-Metzger2016}. \citet{Alexander+2017} detected the electromagnetic counterpart of the binary neutron star merger detected by LIGO/Virgo, GW170817 \citep{Abbott+2017, LIGO+2017} at 6--15 GHz with a flux density of several 100 $\mu$Jy beam$^{-1}$. A superluminal motion of a relativistic jet with flux density of $\sim$50 $\mu$Jy beam$^{-1}$ at 4.5 GHz was subsequently reported by \citet{Mooley+2018}. The expected sensitivity of SKA-VLBI will reach several $\mu$Jy beam$^{-1}$ (see Table \ref{tab:astrometry}) for an integration time of 1 hr at those frequencies, where the physical scale corresponding to $\sim$0.8 mas at 100 Mpc is $\sim$1 pc. Therefore, SKA-VLBI will be able to make movies of expanding ejecta or jets of merging NSNSs or NSBHs. SKA-VLBI is also expected to witness the precession of supermassive black holes (SMBHs), and provide a unique way to obtain a resolved image of the electromagnetic counterpart of a gravitational wave event \citep{Reid+2019baas}.

\textit{Other binaries}. In this section we consider a binary star 1.8 kpc away that is similar to IM Pegasi \citep[IM Peg; HR 8703; see][]{Ransom+2012, Fomalont-Reid2004}. The semimajor axis of the primary star is $\sim$0.1 AU, the orbital period is $\sim$25 days, and the flux density at 8.4 GHz is typically $\sim$10 mJy at a distance of 96.4 pc. In such a case, the astrometric precision would reach about 57 $\mu$as (where the flux density is assumed to be 29 $\mu$Jy, i.e., inversely proportional to the square of the distance), corresponding to an angular radius of $\sim$0.1 AU, which is similar to the assumed semimajor axis of the primary star. Therefore, it is expected the SKA-VLBI will provide high-precision astrometry of binary star orbits, at least for binaries within a distance of $\sim$1.8 kpc. Table 5 in \citet{Schwarz+2016} lists catalogs of binary stars, and \citet{Mason+2001} cataloged over 100,000 binary stars. Radio stars or OH/IR stars can also be sources of binary stars.

\subsubsection{Exoplanets and Dwarfs}\label{sec:science-binary-ED}

The first exoplanets were discovered when precise timing measurements were conducted of the 6.2 ms pulsar PSR1257$+$12 using the Arecibo radio telescope \citep{Wolszczan-Frail1992}. In that work, the pulsar was thought to be orbited by two or more planet-sized bodies with masses of at least 2.8 and 3.4 $M_{\bigoplus}$ (where $M_{\bigoplus}$ is the mass of Earth). Searching for exoplanets using VLBI dates back to the work of \citet{Lestrade+1996}, where observations over a 7.5 yr period excluded a Jupiter-mass planet orbiting at a radius of 4 AU from the central star $\sigma^2$ CrB, with postfit residuals of $<$0.3 mas \citep{Reid-Honma2014}. Subsequently, \citet{Guirado+1997} observed the active K0 star AB Dor and their result indicated a companion with a mass of $\sim$0.1 $M_{\odot}$. Two projects which focused planets and dwarfs, respectively, were subsequently conducted: the Radio Interferometric Planet (RIPL) Search \citep{Bower+2009} and the Radio Interferometric Survey of Active Red Dwarfs \citep[RISARD,][]{Gawronski+2013}. The RIPL collaboration reported on studies of GJ 896A, and found that the upper limit of the companion of GJ 896A was 0.15 $M_J$ (where $M_J$ is the mass of Jupiter) with a semimajor axis of 2 AU, where the precision of the parallax and proper motion were 0.2 mas and 0.01 mas yr$^{-1}$, respectively \citep{Bower+2011}.

Recently, \citet{Curiel+2022} found that the mass and the orbit period are 2.3 $M_J$ and 284.4 days, respectively, for the single planet of the main star (GJ 896A) of the binary, GJ 896AB, through orbit monitoring with VLBA, and optical and infrared data.
However, it is still poorly understood for outer planetary systems \citep{Reid+2019baas}. Fortunately, \citet{Schwarz+2016} cataloged binary or multiple star system containing exoplanets,\footnote{See \url{https://adg.univie.ac.at/schwarz/multiple.html}.} with a total of 217 candidates, and the Extrasolar Planets catalog\footnote{See \url{http://exoplanet.eu/catalog/}.} also lists 4000 planetary systems \citep{Butler+2006, Le-Sidaner+2007, Schneider+2011}. These objects can serve as pilot target sources for investigating exoplanets and potential dwarfs by using SKA-VLBI, which has a higher sensitivity than VLBA.

\subsubsection{Central Black Hole Systems in Galaxies}\label{sec:science-BH-Galaxy}

The nearest SMBH is Sgr A$^{\ast}$ at the center of the MW. Monitoring pulsar orbits around Sgr A$^{\ast}$ with precision astrometry has allowed astronomers to measure and test GR effects \citep[although the observational frequency needs to be greater than about 10 GHz due to interstellar scattering;][]{Reid-Honma2014}. Reports of monitoring observations of pulsars near the GC include only PSR J1745-2900 \citep[see][]{Mori+2013, Bower+2015}. It is an opportunity for SKA-VLBI to conduct orbital monitoring of more pulsars to test GR with high precision. In addition, measuring the proper motion of Sgr A$^{\ast}$ will allow astronomers to constrain the mass of Sgr A$^{\ast}$, evaluate the Sun's orbital motion around the GC, such as its circular rotation speed, $\Theta_0$, and constrain the mass density of SMBH candidates \citep[see][]{Reid+1999, Reid-Honma2014, Reid-Brunthaler2020}.

\subsubsection{Other Multiple Systems}\label{sec:science-binary-Multi-others}

The distances to star clusters can also be determined by using VLBI to measure parallaxes. Although the emission from individual stars in a cluster is very weak, the sensitivity of SKA may allow astronomer to observe them \citep{Fomalont-Reid2004}. At present, the famous and the only example of using VLBI to measure the distance of a star cluster is the Pleiades cluster, where \citet{Melis+2014} measured a distance of 136.2 $\pm$ 1.2 pc, which suggested that the Hipparcos-measured distance to the Pleiades cluster (i.e., 120.2 $\pm$ 1.5 pc) was in error. At the same time, \citet{Melis+2014} proposed that Gaia should identify the unrecognized error in the Hipparcos measurement and correct it. Indeed, Gaia DR2 measured a distance of 135.15 $\pm$ 0.43 pc, which ended the ``Pleiades distance controversy'' \citep{Lodieu+2019}. The flux of member stars in the Pleiades cluster is around several 100 $\mu$Jy at $\sim$8 GHz \citep[see][]{Melis+2014}. Assuming a flux density of 100 $\mu$Jy, SKA-VLBI can also measure the parallax of this cluster with a theoretical precision of $\sim$7$\times 10^{-4}$, i.e., a distance error of $\sim$0.1 pc, where the adopted integration time and bandwidth are shown in Table \ref{tab:astrometry}. The high-sensitivity SKA-VLBI has the potential to measure precisely the parallaxes to many more stellar clusters hosting radio stars (see Section \ref{sec:science-isolate-RS}), and compare them with those determined by Gaia.

From a few detections of 6.7 GHz CH$_3$OH masers in external galaxies (only detected in the Large Magellanic Cloud and M31, and tentatively detected in Maffei 2), it has been found that they are likely located in star-forming regions (clusters) and have peak luminosities in the range $\sim$500--$2.3\times 10^4$ Jy kpc$^2$ \citep{Chen+2022}. Assuming a peak luminosity of $10^4$ Jy kpc$^2$, the peak flux density of a 6.7 GHz CH$_3$OH maser 100 kpc away can reach $\sim$1 Jy. Therefore, it is expected that SKA-VLBI can be used to measure precisely the parallaxes and proper motions of external galaxies within 60 kpc using 6.7 GHz methanol masers as tracers (assuming a parallax precision of $\sim$1.7 $\mu$as, i.e., 1/3 of the astrometric precision shown in Table \ref{tab:astrometry}). If the parallax precision is $\sim$1/5 of astrometric precision (which is achievable), the distances of external galaxies whose parallaxes can be precisely measured can reach even $\sim$100 kpc.

For a more distant galaxy, its distance can be derived by measuring the proper motion of two or more target sources within it, and comparing them with the galaxy's rotation model after obtaining its inclination angle, which is known as the van Maanen experiment \citep{van-Maanen1923, Fomalont-Reid2004, Reid-Honma2014}. The distance to M33 (which is about 730 kpc) was measured using this method \citep{Brunthaler+2002, Brunthaler+2005, Brunthaler+2008}. Two distance galaxies, M31 and Maffei 2 (with distances of $\sim$0.8 and $\sim$5.7 Mpc, respectively), have (tentatively) detected 6.7 GHz CH$_3$OH masers \citep[see][]{Sjouwerman+2010, Chen+2022}.
Therefore, 6.7 GHz CH$_3$OH masers may be a potential tracer to measure distances via the van Maanen experiment. Such measurements of proper motions and geometric distances of galaxies within the Local Group are helpful to understand the Local Group's history, present state and future \citep{Brunthaler+2008}. By applying the measured distances of external galaxies using the van Maanen experiment to distances of standard candles (e.g., Cepheids, Type Ia supernovae, etc.) inside them, the accuracy of the cosmic distance ladder may be improved to 1\% or better \citep{Fomalont-Reid2004}, and thus improve the accuracy of $H_0$.
The proper motions of galaxies can also be used to study galaxy evolution; for example, it may be possible to predict whether will M 31 and its two satellite galaxies M 33 and IC 10 collide with the MW in the future \citep[e.g.,][]{Reid+2019baas}.

\subsection{Astrometry of Complex Systems}\label{sec:science-complex}

\subsubsection{The Structure of the Milky Way}\label{sec:science-complex-MW}

\begin{figure}[!htb]
	\centering
	\includegraphics[width=0.9\textwidth]{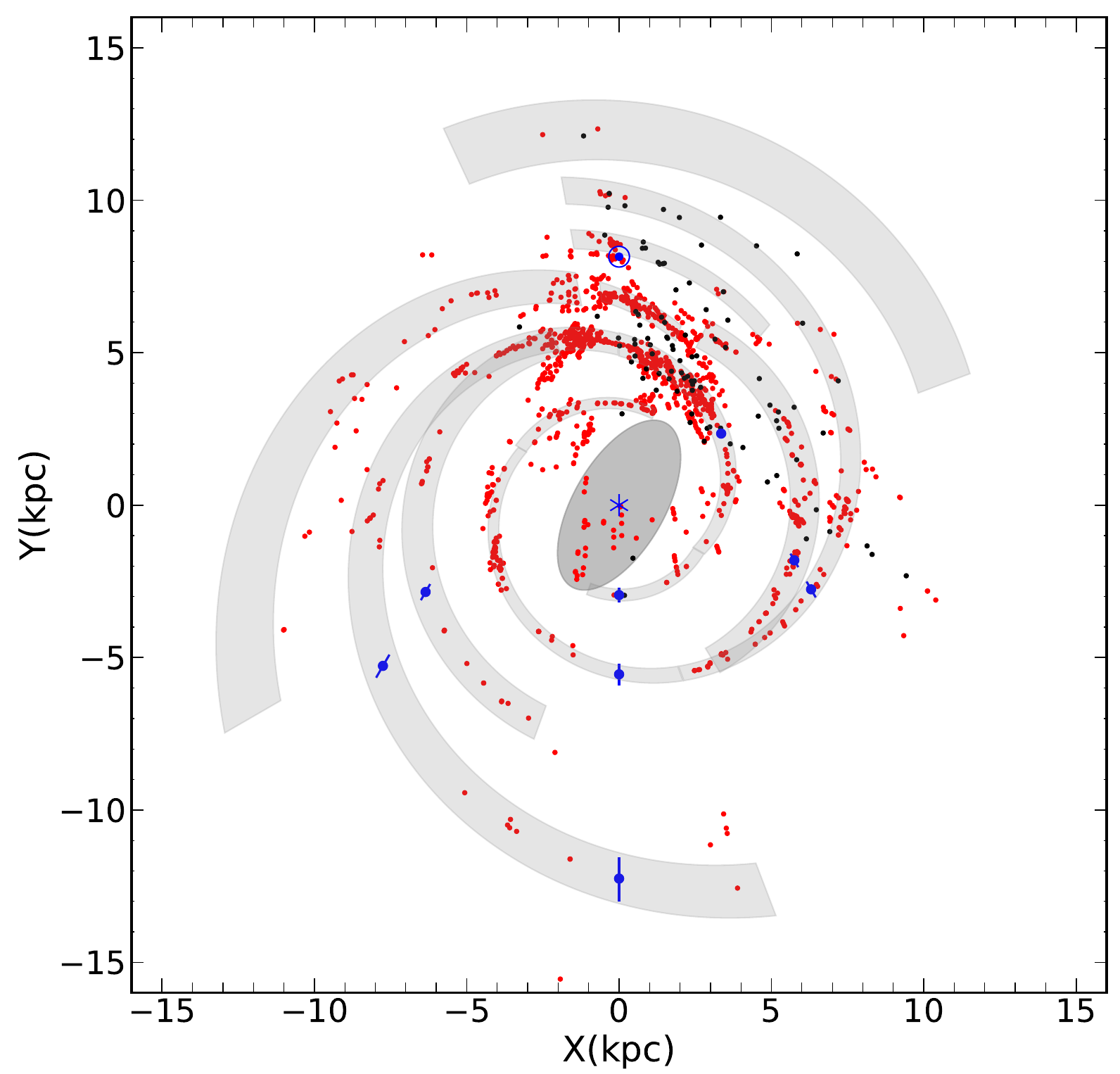}
	\caption{6.7 GHz CH$_3$OH masers superposed on the multiple-arm structure (the gray patterns) of the WM determined by \citet{Xu+2023}, the ellipse indicates the bar following \citet{Hilmi+2020}. The GC (blue star) is at (0, 0) and the Sun (blue Sun symbol) is at (0, 8.15) kpc. The black points indicate the masers from the BeSSeL survey. The red points represent the other masers that can be observed by SKA-VLBI (i.e., Decl. (J2000) $\leq$ 35$^{\circ}$), where the distance is calculated using the kinematic method. The blue dots and error bars show the assumptive sources with the parallax precision of 1.7 $\mu$as.}
\label{fig:spiral}
\end{figure}

Determining the structure of the MW has long been a difficult task. In the past 20 years, advances in astrometry have led to a clearer understanding of the structure of the MW, especially within 5 kpc of the Sun. Such progress is based on high-precision radio astrometry applied to obtain the triangular parallaxes of masers in HMSFRs, and optical Gaia parallaxes of neighboring stars and open clusters \citep{Xu+2018raa, Reid+2019, Hao+2021, Hou2021, Xu+2021}. Nevertheless, to this day, controversy remains regarding such basic facts as the existence of some spiral arms, the number of arms, and the size of the MW. The mainstream view of the structure of the MW is that it is a four-arm spiral galaxy extending from the GC to the outer regions, with some extra arm segments and spurs \citep[see][]{Georgelin-Georgelin1976, Reid+2019}. \citet{Xu+2023} proposed a new paradigm where the MW has a multiple-arm morphology with two symmetrical arms in the inner parts and is split into four spiral arms through two nearly symmetry bifurcations (see Figure \ref{fig:spiral}). This new model suggests the MW may be an ordinary galaxy, as only 2\% of known multiple-arm galaxies have four inner arms, while 83\% of have only two inner arms \citep{Xu+2023}.

The most effective solution to addressing the controversy regarding the spiral structure of the MW is to obtain more parallax measurements of masers in HMSFRs widely distributed in the MW, including in the southern sky and to regions beyond the GC, as well as improving the parallax precision of objects. Compared with Gaia, maser emission is not absorbed significantly by interstellar dust, and is an excellent tracer for the entire MW. SKA-Mid will be located in the Southern Hemisphere. SKA-VLBI can cover regions of the sky with Decl. (J2000) $<$ 35$^{\circ}$ with a theoretical parallax precision up to $\sim$1.7 $\mu$as, which can compensate for the current scarcity of triangular parallax measurements of masers in the southern sky.

Figure \ref{fig:spiral} shows the distribution of 6.7 GHz \choh~masers\footnote{The tracer of the spiral arm structure shown in Figure \ref{fig:spiral} is the 6.7 \choh~maser, because surveys of 12.2 GHz \choh~masers are mostly conducted toward regions where 6.7 GHz \choh~masers were detected, while 22 GHz \ho~masers are not considered in this work.} superposed on the newly determined multiple-arm structure of the WM. The black points indicate 91 masers that currently have a parallax measurement made by the BeSSeL survey \citep{Reid+2019, Xu-Bian-maser+2021, Bian+2022}.
The red points represent masers that can be observed by SKA-VLBI (i.e., Decl. (J2000) $\leq$ 35$^{\circ}$), which include the catalog of \citet{Xu+2009} and several other catalogs (with a large number of newly discovered masers) from 2008 onward, including the southern methanol maser sources observed by ATCA \citep{Caswell2009}, the subsequent Methanol MultiBeam (MMB) survey \citep{Caswell+2010, Green+2010, Caswell+2011, Green+2012, Breen+2015}, the Torun catalogue \citep{Szymczak+2012}, and the survey conducted with the Tianma telescope \citep{Yang+2019}.
There are many masers located near the two bifurcations in the model of \citet{Xu+2023}, which would be a key sample to differentiate between this model and that of \cite{Reid+2019}. To summarize, from this figure, SKA-VLBI is an ideal instrument to solve the controversy over the spiral structure of the MW (see above), and to map the relatively complete structure of the MW.

By comparing the spiral arm structures traced by masers with those traced by OB stars and open clusters obtained from Gaia astrometry, it is possible to study the consistency of the spiral arm structures traced by different young objects and the homogeneity of the distribution of star formation in the spiral arms \citep[e.g.,][]{Xu+2018, Xu+2021}. Naturally, we also expect SKA-VLBI to be able to observe the spiral arm structure depicted by other tracers, such as AGB stars (see Section \ref{sec:science-isolate-PMSS}), which can be compared with the results from Gaia astrometry to study the spiral arm structure traced by old objects \citep[e.g.,][]{Poggio+2021, Lin+2022}. It is also interesting to compare the spiral structure traced by old and young objects \citep[e.g.,][]{Lin+2022}.

The parallaxes and proper motions of masers can also be used to study the structure, motion, and dynamics of spiral arms and the bar \citep[e.g.,][]{Immer+2019, Reid+2019, LiJJ+2022}; the dynamics in the central molecular zone \citep[i.e., the central few 100 pc from the GC; e.g.,][]{Chambers+2014, Kruijssen+2015, Rickert2017} and in the circumnuclear disk \citep[e.g.,][]{Karlsson+2003, Sjouwerman-Pihlstrom2008} of the MW; and to measure the fundamental parameters of the MW. These parameters include $R_0$, $\Theta_0$, the height of the Sun from the Galactic plane, $Z_{\odot}$, and the peculiar motion of the Sun along the direction of its rotation, toward the GC, and the north pole, ($U_{\odot}$, $V_{\odot}$, $W_{\odot}$), \citep[e.g.,][]{Reid+2014, Xu+2018raa, Reid+2019}. Note that $R_0$ is a key parameter that directly affects measurements of the mass and luminosity of almost all objects in the MW \citep{Bland-Hawthorn-Gerhard2016}. In addition, these parallaxes and proper motions can be used to derive the rotation curve of the MW, thus giving kinematic distance models \citep{Reid+2014, Reid+2019, Reid2022}. 3D motion in the vicinity of the solar system can be studied by combining the parallaxes and proper motions of HMSFR masers measured by VLBI and OB stars and young open clusters by Gaia, which produce local velocity field and improve kinematics distance models \citep{Liu+2023}.

\subsubsection{Proper Motion Surveys of Extragalactic Systems}\label{sec:science-complex-AGN-PM}

The expansion or contraction of the universe directly manifests itself as a change in the relative position between pairs of galaxies or galaxy clusters. Therefore, measuring the proper motions of galaxies or galaxy clusters can directly determine whether the universe is expanding or even accelerating, and subsequently the Hubble constant and the distance of galaxies \citep{Darling2013}. For distant galaxies, the ideal target sources are AGNs. By measuring the proper motions of bright radio source pairs, \citet{Darling2013} obtained inclusive $8.3 \pm 14.9$ $\mu$as yr$^{-1}$ for small separation pairs ($<200$ Mpc, $z<$ 0.1, where $z$ is the redshift), which was not constraining enough to determine whether the universe was Hubble expanding or static. However, for large separation pairs (200--1500 Mpc), they obtained no net convergence or divergence for objects with $z<1$, finding $-2.7 \pm 3.7$ $\mu$as yr$^{-1}$, which is consistent with pure Hubble expansion and significantly inconsistent with a static universe. In order to determine further whether the relative proper motions of source pairs at distances $< 200$ Mpc and $z < 0.1$ support Hubble expansion, more source pairs observed with higher astrometric precision and multiple observations of target sources are required. SKA-VLBI astrometry would be helpful to solve this issue.

Other applications of proper motion measurements of galaxies include measuring the secular aberration drift caused by the finite speed of light and the motion of the observer with respect to a light source; secular parallax, which is the motion of the solar system barycenter with respect to the cosmic microwave background (CMB) rest frame; the rotation of the reference frame; the anisotropic expansion of the universe; gravitational waves; large-scale structure; and baryon acoustic oscillation evolution; for more details see \citet{Darling+2018}. Secular aberration drift, aberration proper motion fields, and rotation are tightly tied to the ICRF, and affect the astrometry to all objects \citep{Titov+2011, Truebenbach-Darling-2017, MacMillan+2019, Charlot2020}.

\begin{figure}
	\centering
	\includegraphics[width=0.7\textwidth]{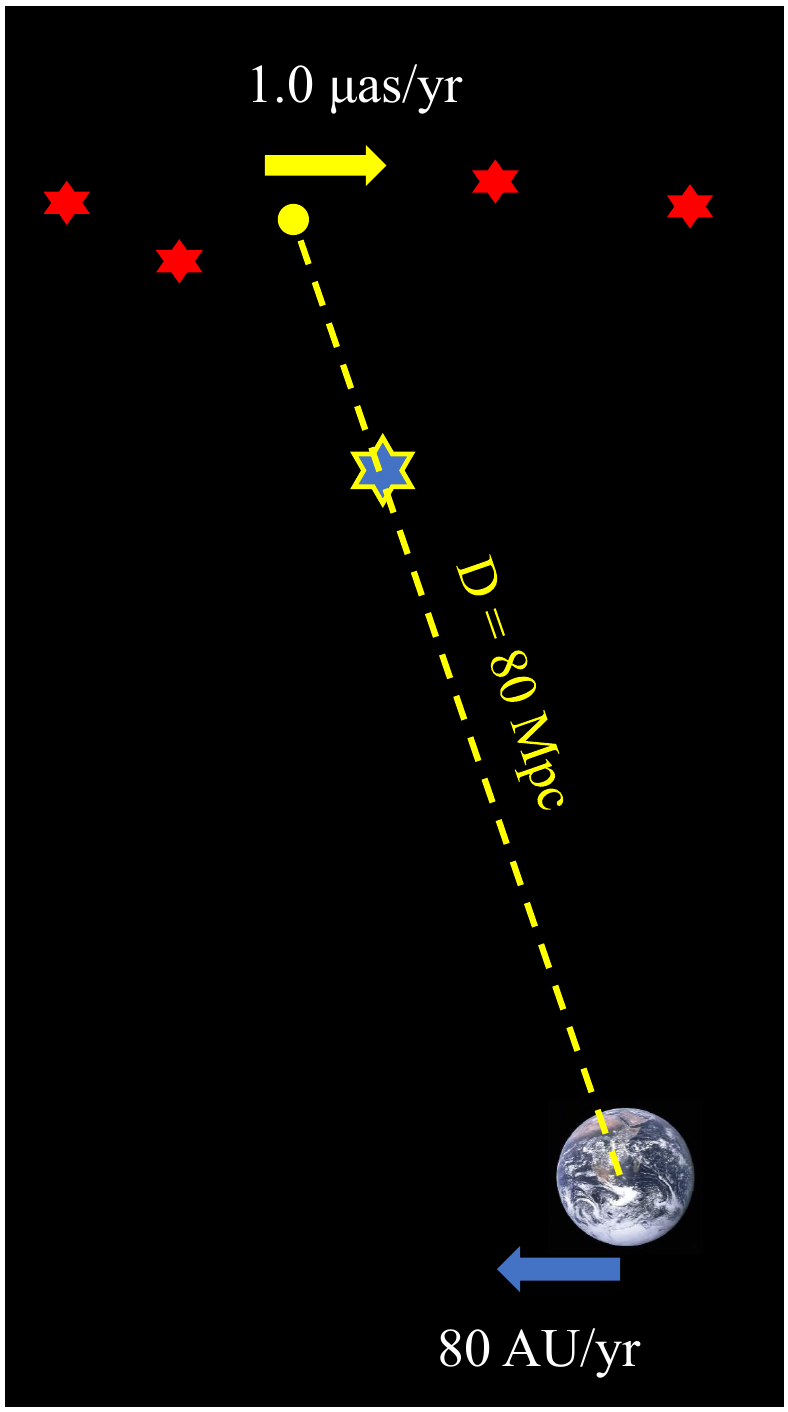}
	\caption{Schematic showing how geometrical distances can be obtained using Earth's motion with respect to the CMB as a baseline. In this model, no motion of a extragalactic object with respect to the CMB and the motion of the Earth of 80 AU yr$^{-1}$ with respect to the center of the CMB is assumed. With a theoretical proper motion precision of 0.2 $\mu$as yr$^{-1}$ achievable through 10 yr of monitoring observations, the distance to an external galaxy 80 Mpc away could be determined with a $\sim$20\% error (corresponding to a proper motion of 1.0 $\mu$as yr$^{-1}$). The icon of Earth is from NASA/Apollo 17 crew, see \url{https://www.nasa.gov/image-article/earth-full-view-from-apollo-17/}.}
	\label{fig:CMB}
\end{figure}

Moreover, the peculiar velocity field on large scales at low redshift is a unique tracer to probe the dark matter density field, where peculiar velocities can be separated from recessional velocities due to the Hubble flow by using directly measured distances \citep{Boruah+2020}. It is feasible to measure the distances to nearby galaxies with SKA-VLBI (see Section \ref{sec:science-binary-Multi-others}), and such measured distance can be used to determine the distances to Cepheids and Type Ia supernova inside those galaxies, and thus improve the precision of the cosmic distance ladder. It is also possible to obtain geometrical distances of extragalactic objects by using a large sample of extragalactic proper motions \citep[][]{Kardashev+1986, Ding+2009}. In this method, it was suggested to use Earth's motion with respect to the CMB, which is $\sim$800 AU over a 10 yr period, as a baseline to measure the proper motions of extragalactic objects. Assuming a positional precision of 2 $\mu$as, 10 yr of monitoring would achieve a theoretical precision of $\sim$0.2 $\mu$as yr$^{-1}$, and a measurable distance with an uncertainty of 20\% could be reached for an object $\sim$80 Mpc distant with a baseline of 800 AU using SKA-VLBI (see Figure \ref{fig:CMB}). Chasing the 1 $\mu$as limit is exciting, as it can lead to measurable distances of galaxies as high as $\sim$160 Mpc.

\subsubsection{Absolute Astrometry and the International Celestial Reference Frame}\label{sec:science-complex-ICRF}

Absolute astrometry allows one to determine the true position of a radio source defined in a well-defined inertial frame \citep{Fomalont-Reid2004}, e.g., the ICRF. The ICRF is a key problem in fundamental astrometry.

Since 1995, the VLBI-based ICRF \citep[i.e., ICRF1,][]{Ma+1998} has been adopted as the IAU standard ICRF, including its second and third versions, ICRF2 and ICRF3, respectively \citep{Fey+2015, Charlot2020}. At the same time, the optical celestial reference frame has also been constantly updated, such as the Hipparcos celestial reference frame (HCRF), and its successor, the GCRF based on Gaia DR2/(E)DR3 \citep{Gaia-Collaboration+2018, Gaia-Collaboration+2022}. Many comparisons between the VLBI-based ICRF and optical celestial reference frame have been made, resulting in the detection of VLBI-Gaia positional offsets (see Section \ref{sec:science-isolate-RS}). Among them, the noise floor of ICRF3 is $\sim$30 $\mu$as for individual source coordinates \citep{Charlot2020}, and the typical positional precision of sources in GCRF is $\sim$1 mas \citep{Gaia-Collaboration+2022}.

\begin{figure}[!htb]
	\centering
	\begin{subfigure}{0.72\textwidth}
		\includegraphics[width=1.0\textwidth]{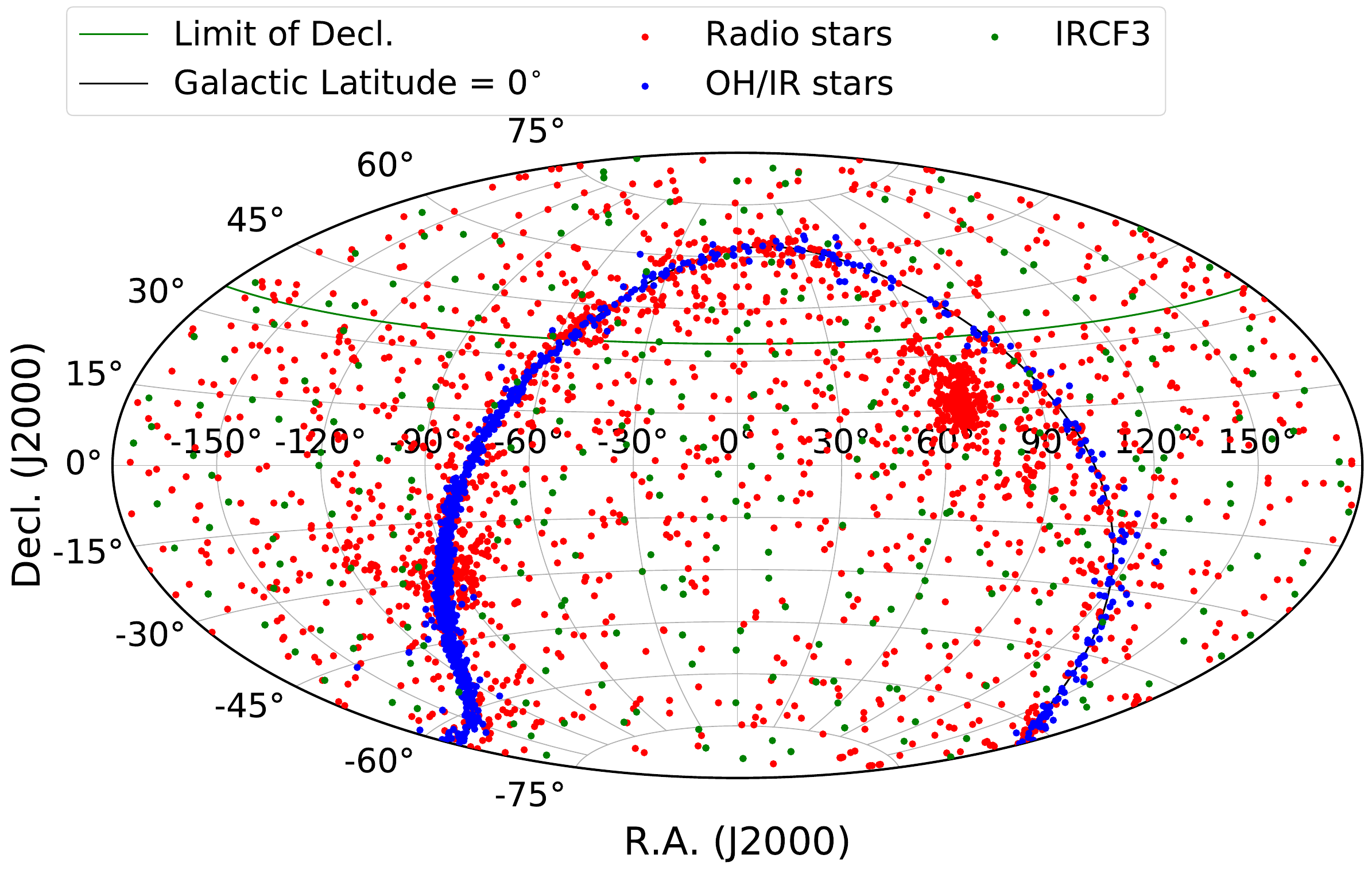}
		\caption{Positional Accuracies from Table \ref{tab:astrometry}.}
	\end{subfigure}
	\begin{subfigure}{0.72\textwidth}
    	\includegraphics[width=1.0\textwidth]{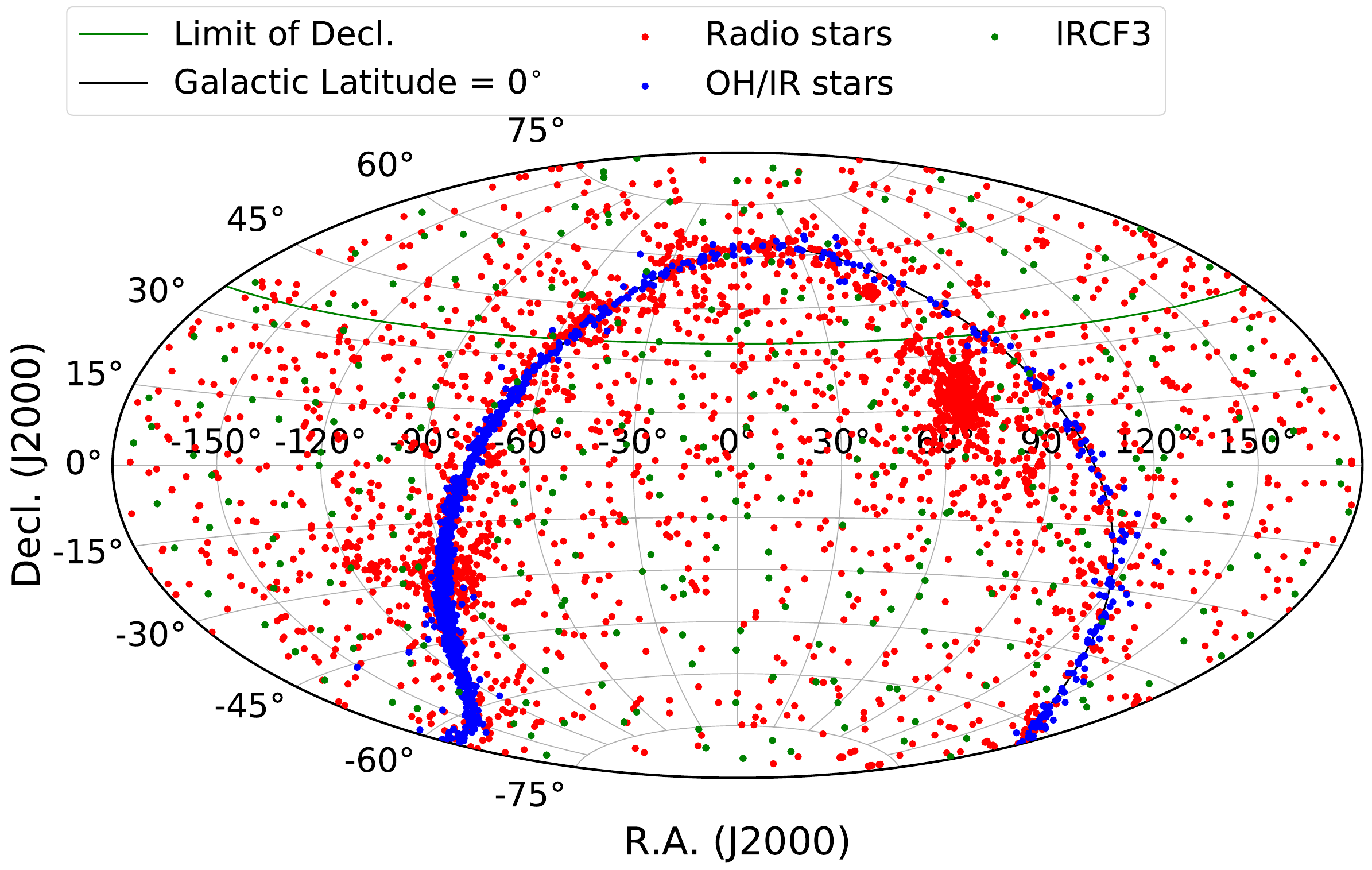}
    	\caption{Positional Accuracies of 30 $\mu$as.}
    \end{subfigure}
	\caption{Distributions of radio stars (red points) and OH/IR stars (blue points) superposed on ICRF3-defining sources (green points). The green and black lines show the Decl. (J2000) limit of SKA-VLBI (i.e., $<$ 35$^{\circ}$) and a Galactic latitude of 0$^{\circ}$, respectively. The criteria of the selected stars are that the flux density (or upper limit) matches the astrometric precision showed in Table \ref{tab:astrometry} (panel (a)), and the astrometric precision is 30 $\mu$as (panel (b)), i.e., the noise floor in the individual source coordinates in ICRF3. For the OH/IR stars, only sources with a distance less than $\sim$16 kpc are shown.
	For the radio stars, only sources that had been observed at 1.0--2.0 and 4.6--15.3 GHz are presented.}
	\label{fig:icrf}
\end{figure}

On the one hand, it is possible to solve the VLBI-Gaia positional offsets by improving the astrometric precision and spatial resolution of radio and optical observations, as well as obtaining more multiband observations and long-term monitoring of ICRF-defining sources. On the other hand, \citet{Froeschle-Kovalevsky1982} proposed using radio stars to link the VLBI-based ICRF with optical celestial reference frames \citep[e.g., HCRF and GCRF; see][]{Kovalevsky+1997, Malkin2016}. This methodology is constantly improving and has achieved fruitful results \citep[e.g.,][]{Malkin2016, Lindegren2020, Bobylev2022, Chen+2023}. We have also presented the distributions of radio stars (see Section \ref{sec:science-isolate-RS}), OH/IR stars (see Section \ref{sec:science-isolate-PMSS}), and ICRF-defining sources in Figure \ref{fig:icrf}, hoping to contribute to future links between GCRF and ICRF using SKA-VLBI. \citet{Makarov+2019} used radio-loud AGNs to link GCRF and ICRF and found that the Gaia DR2 parallax zero-point is color dependent \citep[for the effect, see also][]{Gaia-Collaboration+2021, Lindegren+2021}, suggesting an uncorrected instrumental calibration effect. Linking GCRF and ICRF helps to identify systematic errors in Gaia astrometry which can be corrected in future releases \citep{Rioja-Dodson2020}.

It is possible to determine Earth's rotation and wobbles due to the influence of planets/other objects in the solar system as it orbits the Sun, by measuring the proper motion and/or parallax of Earth relative to an inertial coordinate system \citep[e.g., ICRF;][]{Fomalont-Reid2004}. The proper motion and parallax can also be used to investigate the motion of Earth's crust and its subsequent effect on ground-based telescopes, as well as other important information relevant for geophysical research.

The precision of the ICRF is important for improving the tracking precision of spacecraft, beacons, and satellite orbits and for determining planetary ephemerides. Planetary ephemerides are an important tool, which can be used to calculate the positions of celestial bodies in the solar system at any given time, be incorporated into VLBI data correlators, used for navigation, and employed to predict and/or backtrack astronomical phenomena \citep[e.g.,][]{Li2008}. Determination of modern-day planetary ephemerides requires improving the precision of monitoring planets and their orbiting (artificial) satellites \citep[e.g.,][]{Folkner+2014, Pitjeva-Pitjev2018, Fienga+2020, Park+2021}. Linking planets to the ICRF through VLBI orbital monitoring of artificial satellites can also effectively improve their pointing uncertainties \citep[e.g.,][]{Folkner-Border2015, Park+2021}.

\subsection{Tests of Relativistic Gravity and Gravitational Lenses}\label{sec:science-GR}

\subsubsection{Solar System Tests of Relativistic Gravity}\label{sec:science-GR-solar}

The celestial bodies in the solar system, which define the barycentric celestial reference system (BCRS) and act as gravitational sources, can play an important role in GR tests \citep[for a review see chapter 8 of][]{Ni2017}. There are several ways to conduct a solar system test of GR:
\begin{description}
\item[(1)] Measure light deflection, the Shapiro time delay, and place constraints on the main parameter of parameterized post-Newtonian (PPN) formalism, $\gamma$.
\item[(2)] Measure relativistic perihelion advance and the solar quadrupole moment.
\item[(3)] Conduct ephemerides fitting.
\item[(4)] Measure time variability of the gravitational constant and mass loss from the Sun.
\item[(5)] Quantify frame-dragging effects.
\item[(6)] Conduct lunar laser ranging tests of relativistic gravity.
\end{description}
SKA-VLBI will play an important role in at least points (1)--(5) above. For points (1) and (2), celestial bodies within the solar system act as gravitational sources. The role of SKA-VLBI in determining planetary ephemerides can be found in Section \ref{sec:science-complex-ICRF}, while point (4) is often studied in conjunction with ephemerides fitting. SKA-VLBI can evaluate frame-dragging effects by measuring the proper motions of guide stars \citep[for more details see the satellite mission Gravity Probe B (GP-B);][see also below]{Ratner+2012, Shapiro+2012}.

{\it Light deflection and the Shapiro time delay}. The precision of $\gamma$ obtained by measurements of light deflection and the Shapiro time delay has rapidly improved in recent years \citep[for more details see the reviews by][]{Will2015, Thompson+2017}. VLBI astrometry is an important technique that can lead to improved precision of $\gamma$. The precision of $\gamma$ obtained using the VLBI technique has reached $\sim 9\times10^{-5}$ \citep{Titov+2018} where the Sun acts as a lens. Planets, such as Jupiter, have also been used to test GR using VLBI \citep[e.g.,][]{Fomalont-Kopeikin2003, Fomalont-Kopeikin2008, Li+2022} and data from Gaia EDR3/DR3 \citep[see][]{Abbas+2022}. Light deflection and the underlying theory (e.g., relativistic gravity, gravitational lensing theory, etc.) have become important tools for both astronomy and cosmology \citep{Will2015}, while light deflection by solar system objects will pose a nonnegligible restraining factor in astrometry by SKA-VLBI, affecting positional accuracies at the level $\lesssim$1 $\mu$as \citep{Li-Xu-Bian+2022}.

Taking the solar system planets as gravitational lenses, assuming a positional precision of 2 $\mu$as and there are $\sim$250 sources within 2$^{\circ}$ (see Figure \ref{fig:source-count} at 15 GHz, where the FoV is set to be $\sim$2.1$'$), and that light is deflected by $\sim$2 mas caused by major planets (e.g., Jupiter), it is expected that 10 epochs of observations will produce a measurement precision for $\gamma$ $\sim$10$^{-6}$. If the Sun acts as a lens, the deflection angle of light 1$^{\circ}$ away from the Sun \citep[which is favorable for successful observations; see][]{Titov+2018} can reach $\sim$500 mas, and the number of sources within 1$^{\circ}$ is $\sim$63. Therefore, the measurement precision of $\gamma$ with 10 epochs of observations is expected to reach $\sim$$10^{-8}$, reaching the advanced level of the next few decades \citep[see chapter 8 in][]{Ni2017}. In such a case, using measurements of light deflection caused by the gravitational fields of celestial bodies in the solar system via SKA-VLBI may allow astronomers to test the (noninertial) motion and multipole moment effect of celestial bodies, and test and develop second-order or higher PPN formalisms and/or different metric theories of gravity \citep[e.g.,][]{Damour+Nordtvedt1993, Kopeikin-Makarov2007, Fomalont+2009, Li+2022}. More interestingly, measurements of light deflection may be able to test for the presence of dark energy in the solar system, which requires astrometric precision of $\lesssim$ 10 $\mu$as \citep[see][]{Zhang2022}.

{\it Relativistic perihelion advance}. Another important component of solar system tests of GR is measuring the relativistic perihelion precession. This can be achieved by monitoring the orbits of planets or satellites in the solar system and measuring their annual precession. For example, for Mercury, the contribution of relativistic precession, $\Delta \phi$, is approximately 400 mas yr$^{-1}$ \citep[see page 1113 in][]{Misner+1973, Laskar-Gastineau2009, Mogavero+2021, Brown-Rein2023}. If the positional precision of SKA-VLBI will be 2 $\mu$as, the measurement precision of $\Delta \phi$ with 1 yr of monitoring can reach $\sim 5\times 10^{-6}$. The relationship between $\Delta \phi$ and PPN parameters $\gamma$ and $\beta$ can be expressed by the following equation \citep[see page 1116 in][]{Misner+1973}
\begin{equation}\label{equ:precession}
	\Delta \phi = \left[\frac{(2 - \beta + 2 \gamma)}{3}\right]\left[\frac{6 \pi GM_{\odot}}{c^2 a \left(1 - e^2\right)}\right] + J_2 \left[\frac{3 \pi R_{\odot}^2}{a^2 \left(1 - e^2\right)^2}\right],
\end{equation}
where $G$ is the gravitational constant, $M_{\odot}$ and $R_{\odot}$ are the solar mass and radius, respectively, $a$ and $e$ are the semimajor axis and the eccentricity of the orbit of Mercury, respectively, $c$ is the speed of light, and $J_2$, the solar quadrupole moment, is $\sim 10^{-7}$ \citep[see chapter 8 in][]{Ni2017}. The second term is smaller than the first term by a factor of $\sim 6\times10^{-4}$, and the precision of $J_2$ can currently reach $\sim$10$^{-2}$~\citep[e.g.,][]{Pitjeva+2022}. Therefore, assuming an precision of $\gamma$ of $10^{-6}$, it is expected that SKA-VLBI will allow astronomers to test the PPN parameter $\beta$ at an precision of $\sim$10$^{-6}$. This precision will be comparable to the results obtained by fitting of Ephemerides of Planets and the Moon \citep[EPM; e.g., 3$\times$10$^{-5}$ in][]{Pitjeva2013}, allowing for mutual verification.

{\it Ephemerides fitting and the time derivative of gravitational constant}. As mentioned in Section \ref{sec:science-complex-ICRF}, SKA-VLBI will likely play an important role in improving the precision of ICRF and planetary ephemerides. In fact, while fitting ephemerides, combined with orbital data of (artificial) satellites, GR can also be tested (such as the PPN parameters $\gamma$ and $\beta$ and the solar quadrupole moment, $J_2$); it may even be possible to obtain the time derivative of Newton's gravitational constant \citep[e.g.,][]{Anderson+2002, Fienga+2011, Pitjeva2013, Verma+2014} and whether dark matter present in the solar system \citep[e.g.,][]{Pitjev-Pitjeva2013, Pitjeva-Pitjev2013}; for more details see the review in chapter 8 in \citet{Ni2017}.

{\it Frame-dragging effects}. GP-B was a satellite mission that tested GR frame dragging, i.e., the Lense-Thirring effect, caused by the spin of Earth \citep[see][]{Ratner+2012, Shapiro+2012, Reid-Honma2014}. The task of VLBI in this research was to determine the proper motion of the guide star of the GP-B mission, the RS CVn binary IM Peg, i.e., $-$20.833 $\pm$ 0.090 mas yr$^{-1}$ and $-$27.267 $\pm$ 0.095 mas yr$^{-1}$ in R.A. and Decl. (J2000), respectively. As mentioned previously for IM Peg in Section \ref{sec:science-binary}, SKA-VLBI will be capable of measuring its proper motion with an precision of $\sim$4 $\mu$as yr$^{-1}$, thus improving the precision with which GR frame dragging can be tested. Moreover, the theoretically predicted Lense-Thirring effect signatures of the Galilean moons (such as Io, Europa, Ganymede, and Callisto) can, in principle, currently be detected with an astrometric precision of $\sim$10 mas \citep{Iorio2023}. The flux densities of these satellites were over $\sim$10 mJy, as observed with the VLA at 15 GHz when Jupiter was $\sim$4.46 AU from Earth \citep{de-Pater+1984}. These flux densities are much higher than the sensitivity, $\Delta S$, listed in Table \ref{tab:astrometry} (i.e.,
6.9 $\mu$Jy beam$^{-1}$), indicating SKA-VLBI may be a suitable instrument to measure the Lense-Thirring effect on the Galilean moons of Jupiter.

\subsubsection{Other Tests of the Relativistic Gravity}\label{sec:science-GR-other}

HTBPSRs can be used to test GR (see Section \ref{sec:science-binary}). The measurement of HTBPSR orbital decay depends heavily on the fundamental parameters of the MW (e.g., $R_0$ and $\Theta_0$, see Section \ref{sec:science-complex-MW}), where the acceleration of the Sun and pulsar in their Galactic orbits need to be considered \citep[for the Sun, $\Theta^2_0/R_0$, and for a pulsar, $\Theta^2(R)/R$; for more details see][]{Reid-Honma2014, Deller+2018, Reid+2019baas}. This orbital decay is caused by gravitational radiation, as predicted by GR \citep{Damour-Taylor1991}; thus measuring such decay and deducting the impact of the acceleration of the Sun and pulsar in Galactic orbit provide further tests of GR. A series of studies of the classical HTBPSR, B1913$+$16, showed that VLBA observations of HTBPSRs can test GR with an precision of $\sim 10^{-3}$ \citep{Deller+2018}, but more HTBPSRs and more observations are required (see Section \ref{sec:science-binary}). If we keep the parameters in \citet{Deller+2018} unchanged and simply replace their precision in parallax and proper motion by 6 $\mu$as and 19 $\mu$as yr$^{-1}$ at L2 band, respectively, the calculated ``Galactic'' correction caused by kinematic effects, $\dot{P}^{\mathrm{gal}}_{\mathrm{b}}$, to the observed orbital period derivative is $\dot{P}^{\mathrm{gal}}_{\mathrm{b}} = - (0.008 \pm 0.001) \times 10^{-12}$ s s$^{-1}$, and the ROPR would be 1.0053 $\pm$ 0.0006, as determined by monitoring PSR B1913$+$16. The uncertainties of $\dot{P}^{\mathrm{gal}}_{\mathrm{b}}$ and ROPR are about seven and four times lower than those reported in \citet{Deller+2018}, respectively. It is also expected that more precise values of $R_0$ and $\Theta_0$ will result from higher-precision measurements of the parallaxes and proper motions of densely distributed HMSFRs in the MW (see Section \ref{sec:science-complex-MW}), which may also lead to a more precise measurement of $\dot{P}^{\mathrm{gal}}_{\mathrm{b}}$. Therefore, SKA-VLBI  astrometry will be able to test GR with high precision by observing HTBPSRs. GR has also been tested by monitoring the orbital decay of pulsar binaries \citep[e.g., composed of a pulsar and a white dwarf,][]{Guo+2021} and double neutron star binaries \citep[e.g.,][]{Ding+2021}.

Monitoring pulsars orbit around Sgr A$^\ast$ at the GC can also allow astronomers to measure and test GR effects with high-precision astrometry \citep[see Section \ref{sec:science-BH-Galaxy};][]{Reid-Honma2014}. GR can also be tested by measuring the proper motions of galaxies. The deflection of light from QSOs by a primordial gravitational wave background spanning $10^{-18}$--$10^{-8}$ Hz may also be measured or constrained using SKA-VLBI \citep[see][]{Gwinn+1997, Book-Flanagan2011, Truebenbach-Darling-2017}.

\subsubsection{Gravitational Lenses}\label{sec:science-others-Lens}

Gravitational lenses, a variant of Einstein's light deflection, are an important tool in astronomical research \citep{Will2015}. According to the shapes of images, gravitational lensing can be divided into weak and strong, where the former causes images to be weakly deformed, and the latter results in multiple images \citep[see chapter 8 of][]{Aharonian+2013}. Gravitational microlensing can be thought of as a version of strong gravitational lensing, where the image separation is too small to be resolved \citep[see part 4 in][]{Schneider+2006}.

The application of strong gravitational lensing includes natural telescope that magnifies galaxies or galaxy clusters, enhances the resolution of the instruments, and is beneficial for observing high redshift sources; studies of cosmology; reconstruction of the mass distribution of the lens; etc. Multiple images have the same intrinsic spectrum, structure, and polarization properties. The observed differences between images (such as spectrum and polarization) can be used to study different propagation effects, such as dust extinction and reddening, free-free absorption, and Faraday rotation in the radio regime \citep[for more details see chapter 8 of][]{Aharonian+2013}. Such properties can also be used to measure the Hubble constant \citep[e.g., the series of works by the TDCOSMO collaboration; see][]{Millon+2020}. Gravitational microlensing allows astronomers to study dark energy \citep[see part 4 in ][]{Schneider+2006}, although its prominent application is to search for exoplanets, dwarfs, and dark stars \citep[such as neutron stars, white dwarfs, stellar black holes, etc.;][]{Mao+1991, Zhu-Dong2021, Kaczmarek+2022}. Weak gravitational lensing allows astronomers to reconstruct the mass profile of the lensing galaxy or galaxy cluster, study the mass-luminosity relationship, trace the universal density profile, and study the geometric configuration of the universe \citep[for more details see chapter 8 of][]{Aharonian+2013}.

The advantage of radio VLBI observations of the lensing is that they can provide information at all scales from Mpc to pc, allowing astronomers to study the mass distribution of lenses. However, to date there have been few gravitational lenses detected in radio bands \citep[see chapter 8 of][]{Aharonian+2013}. \citet{Hartley+2021} found signs of the existence of strong- or microlensing in radio images, which helped them study possible radio-quiet QSOs. Radio studies partially assisted by theoretical expectations have presented some possible hints of microlensing, but without clear detections \citep[e.g.,][]{Lecavelier+2013, Vedantham+2020}. Radio research on weak lensing is also beginning to prosper, in the form of several current surveys (e.g., using VLA and/or e-MERLIN) and forthcoming surveys \citep[using SKA; e.g.,][]{Chang+2004, Harrison+2016, Harrison+2020}. However, it is still challenging to identify weak gravitational lensing events using current radio equipment such as VLA $+$ e-MERLIN \citep[e.g.,][]{Harrison+2020}, and there are not many resolved detections \citep{Connor+2022}.

\citet{Koopmans+2004} estimated that SKA will be able to detect $10^6$ strong gravitational lensing events in a half-sky general-purpose survey, of which $10^5$ could be easily detected. \citet{McKean+2015} detailed the observing conditions, i.e., a $\theta_{\mathrm{beam}}$ of 0.25$''$--0.50$''$ and a depth of 3 $\mu$Jy beam$^{-1}$. For Galactic black hole microlensing events, it may be possible to achieve a rate of $\sim$10 yr$^{-1}$ using VLBA with a flux limit of 3 $\mu$Jy beam$^{-1}$ \citep[see][]{Karami+2016}. For exoplanetary microlensing, phase I of SKA-Low may reach a rate of $\sim$0.3 yr$^{-1}$ or better \citep[see][]{Shiohira+2020}. The high-sensitivity SKA will allow for the measurement of weak gravitational lenses in the radio at a level comparable to optical surveys, suggesting its useful application in future weak lensing surveys \citep[see, e.g., chapter 8 of][]{Aharonian+2013, Rivi+2016, Connor+2022}. For example, SKA phase I may be comparable to optical stage III experiments such as the Dark Energy Survey \citep{Bonaldi+2016, Harrison+2016, Camera+2017}. VLBI with SKA-Mid will also be competitive in cosmology studies, especially for $\Lambda$CDM cosmology, where auto-correlations and/or cross-correlation with galaxy weak-lensing and Lyman-$\alpha$ forest observations can be employed \citep[e.g.,][]{Dash+2023}. SKA-VLBI will have a higher resolution compared to SKA, making it suitable for strengthening SKA surveys, and which has great potential to bring about a major breakthrough in gravitational lensing research.

It is worth mentioning that since light deflection caused by a gravitational field is independent of wavelength, gravitational lensing is generally achromatic. However, chromatic microlensing may also be important, e.g., the presence of important chromatic microlensing in strong lensed quasars \citep{Kayser+1986}. One of the important origins of chromatic microlensing may be plasma, while others may be chromatic refractive deflection or selective absorption \citep[for more details see][]{Sun+2023}. Such phenomena, especially plasma lensing, have been studied in the radio band \citep[e.g.,][]{Draine1998, Bisnovatyi-Kogan+2017, Tsupko+2020}, and will therefore be one of the potential subjects of SKA-VLBI.

\section{Summary}\label{sec:sum}

SKA can form VLBI with telescopes on many continents, with baseline lengths mostly at $\sim$8000 km and above, and typical values of $\sim$10,000 km (Section \ref{sec:array-comp}). It is capable of achieving a beam size of several mas or below and an astrometric precision of several $\mu$as (Section \ref{sec:array-sensitivity}) and sensitivity of several $\mu$Jy (Table \ref{tab:astrometry}). The continued development of the AVN will be conducive to forming short and moderate baseline networks linked to SKA-Mid with better uv coverage, and complementing astrometry and mapping observations from dense to extended structures.

The astrometric sciences achievable with SKA-VLBI consists of:

\begin{enumerate}
	
   \item {\it Astrometry of isolated objects (Section \ref{sec:science-isolate}) and binary/multiple systems (Section \ref{sec:science-binary-Multi}), studying objects from star formation to the death of stars}.
   Those objects include protostars (Section \ref{sec:science-isolate-SFR}), main-sequence stars (e.g., star HII 625 in the Pleiades open cluster; Section \ref{sec:science-binary-Multi-others}; \citealt{Melis+2014}), evolved stars like AGB stars (Section \ref{sec:science-isolate-PMSS}), pulsars (Section \ref{sec:science-isolate-PS}), and white dwarfs and black holes (Section \ref{sec:science-binary}). For binary/multiple systems, monitoring their orbital motions may allow astronomers to derive their masses and other physical quantities (Section \ref{sec:science-binary}). Such studies can be extended down to planetary systems (Section \ref{sec:science-binary-ED}) or up to systems surrounding the SMBHs at the centers of galaxies (Section \ref{sec:science-BH-Galaxy}), greatly expanding the study of celestial bodies and celestial systems.
	
   \item {\it Astrometry of complex systems, determining the structure of the MW, refining the ICRF, and enhancing surveys of AGNs or external galaxies.}  To determine better the structure of the MW, observations covering both the Southern Hemisphere and part of the Northern Hemisphere are needed (Section \ref{sec:science-complex-MW}) of objects. Theoretical effective observable distances of up to $\sim$100 kpc allow astrometry (e.g., parallax and proper motion) research to extend to possible nearby dwarf satellite galaxies (Section \ref{sec:science-binary-Multi-others}). Further refinement of the ICRF is needed (Section \ref{sec:science-complex-ICRF}), including studies linking the radio and optical reference frames, perhaps by observing radio stars or radio-loud AGNs. SKA-VLBI will also enhance surveys of AGNs or other external galaxies (Section \ref{sec:science-complex-AGN-PM}), allowing astronomers to investigate the expansion of the universe more precisely, constrain the Hubble constant, and facilitate studies on dark matter.

   \item {\it Geometric distance ladder}. This ladder (which contains at least four components) can be formed from (1) direct trigonometric parallax measurements within the MW (e.g., masers in HMSFRs, Section \ref{sec:science-isolate-SFR}; nearby classic Cepheids, Section \ref{sec:science-isolate-others}; and stellar clusters, Section \ref{sec:science-binary-Multi-others}) and possible neighboring dwarf galaxies within $\sim$100 kpc (Section \ref{sec:science-binary-Multi-others}); (2) galactic distances estimated by possible comparisons of the proper motions of target sources inside galaxies with rotational models (Sections \ref{sec:science-binary-Multi-others}); (3) taking the Earth's motion with respect to the CMB as a baseline to measure the distances to galaxies (by measuring their proper motions; Section \ref{sec:science-complex-AGN-PM}); and (4) using galaxy pair proper motion surveys to measure galactic distances statistically (Section \ref{sec:science-complex-AGN-PM}). This geometric distance technique will cover scales from Earth sized up to a distance of $\sim$80 Mpc or above, providing a coverage range comparable to that from Cepheids to Type Ia supernovae in the cosmic distance ladder. Above all, this geometric distance ladder could provide diverse and independent calibration of the distances of standard candles in the cosmic distance ladder, providing an independent method for determining the Hubble constant.

   \item {\it Fundamental physics, including testing the theory of GR}. SKA-VLBI can be used to test the theory of GR in various ways and from various aspects, such as measuring light deflection caused by the gravitational fields of solar system objects and the perihelion precession of solar system objects (e.g., testing the PPN parameters $\gamma$ and $\beta$ at precision of order $10^{-8}$ and $10^{-6}$, respectively); monitoring the orbits of solar system objects and studying their dynamics (for example, ephemerides fitting); and testing the GR frame drag caused by the rotation of Earth (Section \ref{sec:science-GR-solar}). The by-product of solar system tests of GR can be the study of dark matter and dark energy at these modest scales. Additionally, measuring the parallaxes and proper motions and/or orbital monitoring of HTBPSR; pulsar binaries (with the companion being a white dwarf) or double neutral star binaries; and proper motion surveys of galaxies can also be used to test the theory of GR (see Sections \ref{sec:science-GR-other}).

   \item {\it Synergy with optical astrometry}. Combining optical (e.g., Gaia) and radio astrometry can mutually verify the performance of both (see the example in Section \ref{sec:science-isolate-PMSS}), allowing astronomers to study the structure of the MW (Sections \ref{sec:science-complex-MW}) and refine the ICRF (Section \ref{sec:science-complex-ICRF}).

\end{enumerate}

In summary, in addition to the abundant science goals discussed in this review, we eagerly anticipate hitherto unknown astrometric science with SKA-VLBI, undoubtedly derived from its high sensitivity and high precision.

\normalem
\begin{acknowledgements}
We would like to thank the anonymous referee for the helpful comments and suggestions that helped to improve the paper. This work was funded by the NSFC grants 12203104 and 11933011, the Natural Science Foundation of Jiangsu Province (Grants No. BK20210999), the National SKA Program of China (Grant No. 2022SKA0120103), the Entrepreneurship and Innovation Program of Jiangsu Province, and the Key Laboratory for Radio Astronomy, CAS.
\end{acknowledgements}

\appendix
\section{Telescope Properties and Possible SKA-VLBI Arrays}\label{app:telescope}

A total of 82 telescopes and/or tied arrays are listed here with their array name, e.g., VLBA, EVN (and its subarray MERLIN), EAVN (and its subarrays CVN, KVN, JVN, and VERA), LBA, SKA-Mid (SKA\_M in the figures, and its pathfinders), and the code (which will appear frequently in this work), geographic longitude ($\lambda$) and latitude ($\varphi$), aperture (if it is an array, the number of dishes and the aperture of each dish are presented, separated by ``$\times$''), minimum elevation (EL$_{\mathrm{min}}$), and the SEFD in the L1, L2, C1, C2, C3, X, Ku1, and Ku2 bands, which are centered at 1.3, 1.6, 5.0, 6.0, 6.7, 8.0, 12.0, and 15.0 GHz (i.e., S$_{\mathrm{L1}}$, S$_{\mathrm{L2}}$, S$_{\mathrm{C1}}$, S$_{\mathrm{C2}}$, S$_{\mathrm{C3}}$, S$_{\mathrm{X}}$, S$_{\mathrm{Ku1}}$, and S$_{\mathrm{Ku2}}$), respectively. The corresponding URLs or references or other comments for each telescope are given in the last column.

\setcounter{table}{0}
\renewcommand{\thetable}{A\arabic{table}}

\setcounter{figure}{0}
\renewcommand{\thefigure}{A\arabic{figure}}

\begin{center}
	{
		\tiny
		\setlength\tabcolsep{1.5pt}
		\begin{longtable}{lcllllccccccccccc}
			\caption{Potential Telescopes or Arrays that Could Form a VLBI Array with SKA-Mid\label{tab:telescopes}}\\
			\endfirsthead
			\hline
			\endfoot
			\endlastfoot
			\caption[]{(continued)}\\
			\hline \hline
			Arrays & Index &  Telescope & Code  & Aperture & $\lambda$$^a$ & $\varphi$$^a$ & EL$_{\mathrm{min}}$ & S$_{\mathrm{L1}}$ & S$_{\mathrm{L2}}$ & S$_{\mathrm{C1}}$ & S$_{\mathrm{C2}}$  & S$_{\mathrm{C3}}$ & S$_{\mathrm{X}}$ & S$_{\mathrm{Ku1}}$ & S$_{\mathrm{Ku2}}$ & Ref. \\
			&  &   &   & (m) & ($^{\circ}$) & ($^{\circ}$) & ($^{\circ}$) & (Jy) & (Jy) & (Jy) & (Jy)  & (Jy) & (Jy) & (Jy) & (Jy) &   \\
			\hline
			\endhead
			\hline \hline
			Arrays & Index &  Telescope & Code  & Aperture & $\lambda$$^a$ & $\varphi$$^a$ & EL$_{\mathrm{min}}$ & S$_{\mathrm{L1}}$ & S$_{\mathrm{L2}}$ & S$_{\mathrm{C1}}$ & S$_{\mathrm{C2}}$  & S$_{\mathrm{C3}}$ & S$_{\mathrm{X}}$ & S$_{\mathrm{Ku1}}$ & S$_{\mathrm{Ku2}}$ & Ref. \\
			&  &   &   & (m) & ($^{\circ}$) & ($^{\circ}$) & ($^{\circ}$) & (Jy) & (Jy) & (Jy) & (Jy)  & (Jy) & (Jy) & (Jy) & (Jy) &   \\
			\hline
			VLBA	&	1	&	Saint Croix	&	VLBA\_SC	&	25	&	$-$64.58 	&	18.34 	&	0	&	289	&	314	&	210	&	278	&	278	&	327	&	543	&	543	& 1	\\
			VLBA	&	2	&	Hancock	&	VLBA\_HN	&	25	&	$-$71.99 	&	43.07 	&	0	&	289	&	314	&	210	&	278	&	278	&	327	&	543	&	543	&	1		\\
			VLBA	&	3	&	North Liberty	&	VLBA\_NL	&	25	&	$-$91.57 	&	42.31 	&	0	&	289	&	314	&	210	&	278	&	278	&	327	&	543	&	543	&	1		\\
			VLBA	&	4	&	Fort Davis	&	VLBA\_FD	&	25	&	$-$103.94 	&	31.58 	&	0	&	289	&	314	&	210	&	278	&	278	&	327	&	543	&	543	&	1		\\
			VLBA	&	5	&	Los Alamos	&	VLBA\_LA	&	25	&	$-$106.25 	&	36.31 	&	0	&	289	&	314	&	210	&	278	&	278	&	327	&	543	&	543	&	1		\\
			VLBA	&	6	&	Pie Town	&	VLBA\_PT	&	25	&	$-$108.12 	&	37.33 	&	0	&	289	&	314	&	210	&	278	&	278	&	327	&	543	&	543	&	1		\\
			VLBA	&	7	&	Kitt Peak	&	VLBA\_KP	&	25	&	$-$111.61 	&	32.06 	&	0	&	289	&	314	&	210	&	278	&	278	&	327	&	543	&	543	&	1		\\
			VLBA	&	8	&	Owens Valley	&	VLBA\_OV	&	25	&	$-$118.28 	&	41.63 	&	0	&	289	&	314	&	210	&	278	&	278	&	327	&	543	&	543	&	1		\\
			VLBA	&	9	&	Brewster	&	VLBA\_BR	&	25	&	$-$119.68 	&	56.59 	&	0	&	289	&	314	&	210	&	278	&	278	&	327	&	543	&	543	&	1		\\
			VLBA	&	10	&	Mauna Kea	&	VLBA\_MK	&	25	&	$-$155.46 	&	20.25 	&	0	&	289	&	314	&	210	&	278	&	278	&	327	&	543	&	543	& 1			\\
			VLA	&	11	&	phased VLA	&	VLA27	&	115	&	$-$107.62 	&	34.08 	&	8	&	16	&	14	&	12	&	12	&	12	&	10	&	12	&	12	&	2	\\
			GBT	&	12	&	GBT	&	GBT	&	110	&	$-$79.84 	&	38.43 	&	5	&	10	&	10	&	10	&	10	&	10	&	15	&	20	&	20	&	3	\\
			EVN/MERLIN	&	13	&	Jodrell Bank-1	&	JOD1	&	76	&	$-$2.31 	&	53.24 	&	$-$1	&	35	&	40	&	70	&	80	&	80	&	...	&	...	&	...	&	4	\\
			EVN/MERLIN	&	14	&	Jodrell Bank-2	&	JOD2	&	38$\times$25	&	$-$2.31 	&	53.24 	&	0	&	9	&	5	&	8	&	8	&	8	&	...	&	...	&	...	&	4	\\
			EVN	&	15	&	Westerbork	&	WB\_T	&	14$\times$25	&	6.60 	&	52.92 	&	0	&	31	&	42	&	125	&	119	&	119	&	125	&	...	&	...	&		4, 5 \\
			EVN	&	16	&	Effelsberg	&	EFS	&	100	&	6.88 	&	50.52 	&	8	&	20	&	19	&	20	&	25	&	25	&	20	&	78	&	79	&		6	\\
			EVN	&	17	&	Medicina	&	MEDI	&	32	&	11.65 	&	44.52 	&	5	&	490	&	700	&	170	&	840	&	840	&	320	&	...	&	...	&		4	\\
			EVN	&	18	&	Noto	&	NOTO	&	32	&	14.99 	&	36.88 	&	5	&	740	&	740	&	260	&	1100	&	1100	&	840	&	...	&	...	&		4	\\
			EVN	&	19	&	Sardinia	&	SARD	&	65	&	9.25 	&	39.49 	&	5	&	67	&	67	&	...	&	50	&	50	&	...	&	...	&	...	&		4	\\
			EVN	&	20	&	Onsala	&	ONSA85	&	25	&	11.92 	&	57.39 	&	7	&	310	&	310	&	480	&	850	&	850	&	...	&	...	&	...	&		4	\\
			EVN	&	21	&	Onsala-60	&	ONSA60	&	20	&	11.93 	&	57.40 	&	6	&	...	&	...	&	...	&	...	&	...	&	785	&	...	&	...	&		4	\\
			EVN	&	22	&	Torun	&	TORU	&	32	&	18.56 	&	52.91 	&	2	&	250	&	300	&	220	&	650	&	650	&	...	&	...	&	...	&		4	\\
			EVN	&	23	&	Yebes	&	YEBE	&	40	&	$-$3.09 	&	40.52 	&	5	&	...	&	...	&	160	&	160	&	160	&	200	&	...	&	...	&		4	\\
			EVN	&	24	&	Irbene	&	IRBE	&	32	&	21.85 	&	57.55 	&	3	&	...	&	715	&	285	&	285	&	285	&	450	&	...	&	...	&	4, 7	\\
			EVN	&	25	&		&	IRBE16	&	16	&	21.85 	&	57.56 	&	3	&	...	&	...	&	623	&	558	&	569	&	651	&	...	&	...	&	4, 8	\\
			EVN	&	26	&	Mets\"{a}hovi	&	METS	&	14	&	24.39 	&	60.22 	&	0	&	...	&	...	&	...	&	...	&	...	&	3200	&	...	&	...	&	4	\\
			EVN	&	27	&	Wettzell	&	WETT	&	20	&	12.88 	&	49.15 	&	0	&	...	&	...	&	...	&	...	&	...	&	750	&	...	&	...	&	4	\\
			EVN	&	28	&	Svetloe	&	SVET	&	32	&	29.78 	&	60.53 	&	$-$5	&	360	&	360	&	250	&	...	&	...	&	200	&	...	&	...	&	4	\\
			EVN	&	29	&	Zelenchukskaya	&	ZELE	&	32	&	41.57 	&	43.79 	&	3	&	300	&	300	&	330	&	...	&	...	&	200	&	...	&	...	&	4	\\
			EVN	&	30	&	Badary	&	BADA	&	32	&	102.23 	&	51.77 	&	$-$5	&	330	&	330	&	330	&	...	&	...	&	200	&	...	&	...	&	4	\\
			EVN/MERLIN	&	31	&	Cambridge	&	CAMB	&	32	&	0.04 	&	52.17 	&	2	&	200	&	175	&	175	&	225	&	225	&	...	&	...	&	...	&	4	\\
			EVN/MERLIN	&	32	&	Darnhall	&	MERLI1/DARN	&	25	&	2.54 	&	53.16 	&	2	&	245$^b$	&	250	&	188	&	188	&	188	&	...	&	...	&	...	&	4 	\\
			EVN/MERLIN	&	33	&	Defford	&	DEFF	&	25	&	2.14 	&	52.10 	&	2	&	350	&	350	&	1000	&	1600	&	1600	&	...	&	...	&	...	&	4	\\
			EVN/MERLIN	&	34	&	Knockin	&	MERLI1/KNOC	&	25	&	3.00 	&	52.79 	&	2	&	$-$$^b$	&	$-$	&	$-$	&	$-$	&	$-$	&	...	&	...	&	...	&	4	\\
			EVN/MERLIN	&	35	&	Pickmere	&	MERLI1/TABL	&	25	&	2.45 	&	53.29 	&	2	&	$-$$^b$	&	$-$	&	$-$	&	$-$	&	$-$	&	...	&	...	&	...	&	4	\\
			EVN	&	36	&	Robledo/DSS63	&	DSS63	&	70	&	$-$4.25 	&	40.43 	&	6	&	...	&	27	&	...	&	...	&	...	&	18	&	...	&	...	&	4	\\
			EVN	&	37	&	DSS65	&	DSSS	&	34	&	$-$4.25 	&	40.43 	&	6	&	...	&	...	&	...	&	...	&	...	&	61$^c$	&	...	&	...	&	4	\\
			EVN	&	38	&	DSS-54	&	DSSS	&	34	&	$-$4.25 	&	40.43 	&	6	&	...	&	...	&	...	&	...	&	...	&	$-$$^c$	&	...	&	...	&	4	\\
			EVN	&	39	&	DSS-55	&	DSSS	&	34	&	$-$4.25 	&	40.43 	&	6	&	...	&	...	&	...	&	...	&	...	&	$-$$^c$	&	...	&	...	&	4	\\
			EVN	&	40	&	Matera	&	MATE	&	20	&	16.70 	&	40.65 	&	5	&	...	&	...	&	...	&	...	&	...	&	900	&	...	&	...	&	9	\\
			EVN	&	41	&	Ny-Alesund	&	NYAL	&	20	&	11.87 	&	78.93 	&	2	&	...	&	...	&	...	&	...	&	...	&	1255	&	...	&	...	&	4	\\
			EVN/CVN/EAVN	&	42	&	Sheshan	&	SESHAN	&	25	&	121.14 	&	31.09 	&	8	&	...	&	670	&	720	&	1500	&	1740	&	800	&	...	&	...	&	4, 10	\\
			EVN/CVN/EAVN	&	43	&	Tianma	&	TIANMA	&	65	&	121.20 	&	31.10 	&	5	&	39	&	46	&	26	&	26	&	55	&	48	&	56	&	56	&	4, 11	\\
			EVN/CVN/EAVN	&	44	&	Nanshan	&	NANSH	&	26	&	87.18 	&	43.47 	&	5	&	300	&	300	&	250	&	...	&	...	&	350	&	...	&	...	&	4, 10	\\
			EVN/CVN/EAVN	&	45	&	Kunming	&	KUMI	&	40	&	102.80 	&	25.03 	&	8	&	...	&	...	&	350	&	290	&	307	&	480	&	...	&	...	&	4, 10	\\
			EVN/KVN/EAVN	&	46	&	KVN Yonsei	&	YSKVN	&	21	&	126.94 	&	37.57 	&	5	&	...	&	...	&	...	&	...	&	4241	&	...	&	...	&	...	&	4, 10	\\
			EVN/KVN/EAVN	&	47	&	KVN Ulsan	&	KUKVN	&	21	&	129.25 	&	35.55 	&	5	&	...	&	...	&	...	&	...	&	4241	&	...	&	...	&	...	&	4, 10	\\
			EVN/KVN/EAVN	&	48	&	KVN Tamna	&	KTKVN	&	21	&	126.46 	&	33.29 	&	5	&	...	&	...	&	...	&	...	&	4241	&	...	&	...	&	...	&	4, 10	\\
			EVN/AVN/LBA	&	49	&	Hartebeesthoek	&	HART26	&	26	&  	27.69 	&	$-$25.89 	&	10	&	...	&	430$^d$	&	650	&	700	&	700	&	630	&	1175	&	...	&	4, 12	\\
			EVN/AVN/LBA	&	50	&		&	HART15	&	15	&	27.67 	&	$-$25.89 	&	10	&	...	&	...	&	...	&	...	&	...	&	1400	&	...	&	...	&	4, 12	\\
			EAVN	&	51	&	Nobeyama	&	NOBE	&	45	&	138.47 	&	35.94 	&	12	&	...	&	...	&	...	&	...	&	...	&	...	&	...	&	...	&	10	\\
			EAVN	&	52	&	Takahagi	&	TAKA	&	32	&	140.69 	&	36.70 	&	5	&	...	&	...	&	...	&	...	&	200	&	...	&	...	&	...	&	10	\\
			EAVN	&	53	&	Hitachi	&	HITA	&	32	&	140.69 	&	36.70 	&	5	&	...	&	...	&	...	&	...	&	200	&	...	&	...	&	...	&	10	\\
			EAVN	&	54	&	Yamaguchi	&	YAMA	&	32	&	131.56 	&	34.22 	&	5	&	...	&	...	&	...	&	...	&	290	&	...	&	...	&	...	&	10	\\
			EAVN	&	55	&	Sejong	&	SEJO	&	22	&	127.30 	&	36.52 	&	0	&	...	&	...	&	...	&	...	&	...	&	...	&	...	&	...	&	10	\\
			EAVN/VERA/JVN	&	56	&	Mizusawa	&	MIZU	&	20	&	141.13 	&	39.13 	&	5	&	...	&	...	&	...	&	...	&	2300	&	...	&	...	&	...	&	10	\\
			EAVN/VERA/JVN	&	57	&	Iriki	&	IRIK	&	20	&	130.44 	&	31.75 	&	5	&	...	&	...	&	...	&	...	&	2300	&	...	&	...	&	...	&	10	\\
			EAVN/VERA/JVN	&	58	&	Ogasawara	&	OGAS	&	20	&	142.22 	&	27.09 	&	5	&	...	&	...	&	...	&	...	&	2300	&	...	&	...	&	...	&	10	\\
			EAVN/VERA/JVN	&	59	&	Ishigakijima	&	ISHI	&	20	&	124.17 	&   11	24.41 	&	5	&	...	&	...	&	...	&	...	&	2300	&	...	&	...	&	...	&	10	\\
			EAVN/JVN	&	60	&	Gifu	&	GIFU	&	11	&	136.74 	&	35.47 	&	10	&	...	&	...	&	...	&	...	&	...	&	...	&	...	&	...	&	13, 14	\\
			EAVN/JVN	&	61	&	Usuda	&	USUD	&	64	&	138.36 	&	36.13 	&	7	&	...	&	...	&	...	&	...	&	137	&	...	&	...	&	...	&	13, 14	\\
			LBA	&	62	&	ATCA/Narrabri	&	ATCA	&	6$\times$22	&	149.57 	&	$-$30.31 	&	12	&	57	&	57	&	58	&	58	&	142	&	72	&	217	&	...	&	15	\\
			LBA	&	63	&	Mopra	&	MOPRA	&	22	&	149.07 	&	$-$31.30 	&	12	&	340	&	340	&	350	&	350	&	850	&	430	&	1300	&	...	&	15	\\
			LBA	&	64	&	Parkes	&	PARKS	&	64	&	148.26 	&	$-$33.00 	&	30	&	40	&	42	&	110	&	110	&	110	&	43	&	370	&	...	&	15	\\
			LBA	&	65	&	Tidbinbilla/DSS43	&	DSS43	&	70	&	148.98 	&	$-$35.40 	&	6	&	...	&	23	&	...	&	...	&	...	&	25	&	...	&	....	&	15	\\
			LBA	&	66	&	Tidbinbilla/DSS45	&	DSS45	&	34	&	148.98 	&	$-$35.40 	&	8	&	...	&	...	&	...	&	...	&	...	&	87	&	...	&	...	&	15	\\
			LBA	&	67	&	Hobart	&	HOBART26	&	26	&	147.44 	&	$-$42.80 	&	16	&	470	&	420	&	640	&	...	&	1240	&	560	&	1200	&	...	&	15	\\
			LBA	&	68	&		&	HOBART12	&	12	&	147.44 	&	$-$42.80 	&	10	&	...	&	...	&	3500	&	3500	&	3500	&	3400	&	3400	&	...	&	16, 17	\\
			LBA	&	69	&	Ceduna	&	CEDUNA	&	12	&	133.81 	&	$-$31.87 	&	20	&	...	&	...	&	450	&	...	&	550	&	600	&	750	&	...	&	15	\\
			LBA	&	70	&	Katherine	&	KATH	&	12	&	132.15 	&	$-$14.40 	&	10	&	...	&	...	&	3500	&	3500	&	3500	&	3400	&	3400	&	...	&	16, 17	\\
			LBA	&	71	&	Yarragadee	&	YARR	&	12	&	115.35 	&	$-$29.00 	&	10	&	...	&	...	&	3500	&	3500	&	3500	&	3400	&	3400	&	...	&	16, 17	\\
			LBA	&	72	&	Warkworth	&	WARK12	&	12	&	174.66 	&	$-$36.40 	&	7	&	...	&	...	&	3500	&	3500	&	3500	&	3400	&	3400	&	...	&  	18--20	\\
			LBA	&	73	&		&	WARK30	&	30	&	174.66 	&	$-$36.40 	&	6	&	...	&	...	&	...	&	...	&	650	&	...	&	...	&	...	&	18--20	\\
			LBA/SKA	&	74	&	ASKAP	&	ASKAP	&	36$\times$12	&	116.63 	&	$-$26.70 	&	15	&	70	&	80	&	...	&	...	&	...	&	...	&	...	&	...	& 21, 22		\\
			AVN/SKA	&	75	&	MeerKAT	&	MEER	&	64$\times$13.5	&	21.44 	&	$-$30.71 	&	16	&	9$^e$	&	...	&	...	&	...	&	...	&	...	&	...	&	...	&	23	\\
			AVN/SKA	&	76	&	SKA-Mid	&	SKA\_M	&	133$\times$15	&	21.44 	&	$-$30.71 	&	15	&	2.6$^f$	&	2.6$^f$	&	2.6$^f$	&	2.6$^f$	&	2.6$^f$	&	2.6$^f$	&	3.9$^f$	&	3.9$^f$	&	24	\\
			AVN	&	77	&	Ghana	&	GHAN	&	32	&	$-$0.31 	&	5.75 	&	0	&	...	&	...	&	563	&	...	&	563	&	...	&	...	&	...	&	24--27	\\
			CVN	&	78	&	Miyun	&	MIYU	&	50	&	116.98 	&	40.56 	&	5	&	...	&	...	&	...	&	...	&	...	&	227	&	...	&	...	&	28	\\
			... &	79	&	FAST	&	FAST	&	500	&	106.86 	&	25.65 	&	50	&	2	&	...	&	...	&	...	&	...	&	...	&	...	&	...	&	29$-$30	\\
			...	&	80	&	uGMRT	&	GMRT	&	30$\times$45	&	74.05 	&	19.10 	&	16	&	10	&	10	&	...	&	...	&	...	&	...	&	...	&	...	&	31	\\
			... &	81	&	TNRO	&	TNRO	&	40	&	99.22 	&	18.86 	&	10	&	78	&	78	&	...	&	...	&	...	&	...	&	...	&	...	&	32\\
			... &	82	&	Jingdong	&	JING	&	120	&	101.00 	&	24.50 	&	5	&	6	&	6	&	6	&	6	&	6	&	...	&	...	&	...	&	33	\\
			\hline
\multicolumn{17}{l}{\parbox{\textwidth}{Notes. $a$. For telescopes in Europe, see \url{https://www.craf.eu/radio-observatories-in-europe/}, for others see column ``Ref."}} \\
\multicolumn{17}{l}{\parbox{\textwidth}{$b$. We have represented the three telescopes in MERLIN as MERLIN1 because they are close to each other and their performances are similar.}} \\
\multicolumn{17}{l}{\parbox{\textwidth}{$c$. We have represented the three telescope as DSSS because they are close to each other and their performances are similar.}}\\
\multicolumn{17}{l}{\parbox{\textwidth}{$d$. The minimum frequency is 1608 MHz.}} \\
\multicolumn{17}{l}{\parbox{\textwidth}{$e$. We only considered its L band, with frequency between 900 and 1670 MHz.}}\\
\multicolumn{17}{l}{\parbox{\textwidth}{\begin{spacing}{0.5}$f$. Only consider the inner 4 km core \citep{Paragi+2015}. We assume that all telescopes equip with receivers of all bands, but initially only a fraction of the telescopes will be equipped with Band 5 (i.e., 8300--15,300 MHz) receivers.\end{spacing}}}\\[0.25cm]
\multicolumn{17}{l}{\parbox{\textwidth}{References: (1) \url{https://science.nrao.edu/facilities/vlba};  (2) \url{https://science.nrao.edu/facilities/vla/docs};  (3)  \url{https://www.gb.nrao.edu/scienceDocs};  (4) \url{http://old.evlbi.org/user_guide};  (5) \citet{van-Cappellen+2022};  (6) \url{https://eff100mwiki.mpifr-bonn.mpg.de/doku.php};  (7) \url{https://www.virac.eu/en/research/technical-parameters};  (8) \citet{Bleiders+2017};  (9) \citet{Colucci+1999};  (10) \url{https://radio.kasi.re.kr/eavn/main.php};  (11) \url{http://radio-en.shao.cas.cn/facility/};  (12) \url{http://www.hartrao.ac.za/index.php};  (13) \url{http://astro.sci.yamaguchi-u.ac.jp/jvn/eng/status_e.html};  (14) \citet{Wajima+2016};  (15) \url{https://www.atnf.csiro.au/vlbi/};  (16) \citet{Lovell+2013};  (17) \citet{Hyland+2023};  (18) \citet{Woodburn+2015};  (19) \citet{Petrov+2015};  (20) \citet{Weston+2010};  (21) \url{https://www.atnf.csiro.au/projects/askap/index.html};  (22) \citet{Hotan+2021};  (23) \url{https://skaafrica.atlassian.net/wiki/spaces/ESDKB/pages/277315585/MeerKAT+specifications};  (24) \url{https://www.skao.int/en/science-users/}; \citet{Paragi+2015} (24) \citet{Duah-Asabere+2015};  (25) \citet{de-Witt+2016};  (26) \citet{Azankpo2017};  (27) \citet{Venter-Malan2021};  (28) \citet{Zhang+2009};  (29) \citet{Jiang+2019};  (30) \citet{Jiang+2020};  (31) \url{http://www.ncra.tifr.res.in/ncra/gmrt};  (32) \citet{Jaroenjittichai+2022};  (33) \citet{Wang+2022}.}}	
\\
		\end{longtable}
	}
\end{center}

\bibliographystyle{raa}
\bibliography{liyj}
\end{CJK*}
\end{document}